\documentclass[acmsmall]{acmart}

\usepackage{csquotes}
\usepackage[export]{adjustbox}
\usepackage{caption}
\usepackage{subcaption}
\usepackage{tabularx}
\usepackage{amsthm}
\usepackage{amsmath} 
\usepackage{ragged2e}
\usepackage{extarrows}
\usepackage{pifont}
\usepackage{algorithm2e}
\usepackage{algpseudocode}

\newtheorem{Assumption}{Assumption}
\newtheorem{Proposition}{Proposition}
\usepackage{url} 

\def\BibTeX{{\rm B\kern-.05em{\sc i\kern-.025em b}\kern-.08emT\kern-.1667em\lower.7ex\hbox{E}\kern-.125emX}}
    
\setcopyright{acmlicensed}
\acmYear{2022 Accepted by TOIS}

\begin{document}

%
\title{Towards Equivalent Transformation of User Preferences in Cross Domain Recommendation}

\author{Xu Chen}
\affiliation{%
  \institution{Shanghai Jiao Tong University and University of Technology, Sydney}
  \streetaddress{Dongchuan Rd.800; 15 Broadway Broadway NSW}
  \city{Shanghai; Sydney}
  \country{China; Australia}}
\email{xuchen2016@sjtu.edu.cn}

\author{Ya Zhang}
\authornote{Prof. Ya Zhang is the corresponding author.}
\affiliation{%
  \institution{Shanghai Jiao Tong University}
  \city{Shanghai}
  \country{China}}
\email{ya\_zhang@sjtu.edu.cn}

\author{Ivor W. Tsang}
\affiliation{%
  \institution{Shanghai Jiao Tong University}
  \city{Sydney}
  \country{Australia}}
\email{ivor.tsang@uts.edu.au}

\author{Yuangang Pan}
\affiliation{%
  \institution{University of Technology, Sydney}
  \city{Sydney}
  \country{Australia}}
\email{Yuangang.pan@uts.edu.au}

\author{Jingchao Su}
\affiliation{%
  \institution{Shanghai Jiao Tong University}
  \city{Shanghai}
  \country{China}}
\email{sujingchao@sjtu.edu.cn}

\renewcommand{\shortauthors}{Xu Chen et al.}

%
\begin{abstract}
Cross domain recommendation (CDR) is one popular research topic in recommender systems. This paper focuses on a popular scenario for CDR where different domains share the same set of users but no overlapping items. The majority of recent methods have explored the shared-user representation to transfer knowledge across domains. 
However, the idea of shared-user representation resorts to learn the overlapped features of user preferences and suppresses the domain-specific features. Other works try to capture the domain-specific features by an MLP mapping but require heuristic human knowledge of choosing samples to train the mapping. In this paper, we attempt to learn both features of user preferences in a more principled way. We assume that each user's preferences in one domain can be expressed by the other one, and these preferences can be mutually converted to each other with the so-called equivalent transformation. Based on this assumption, we propose an equivalent transformation learner (ETL) which models the joint distribution of user behaviors across domains. The equivalent transformation in ETL relaxes the idea of shared-user representation and allows the learned preferences in different domains to preserve the domain-specific features as well as the overlapped features. Extensive experiments on three public benchmarks demonstrate the effectiveness of ETL compared with recent state-of-the-art methods. Codes and data are available online:~\url{https://github.com/xuChenSJTU/ETL-master}
\end{abstract}



\maketitle

\section{Introduction}
Cross domain recommendation (CDR)~\cite{cantador2015cross} has been proposed to solve the data sparsity problem in recommendation by transferring knowledge from other domains. According to real-world application conditions, CDR can be categorized into four different scenarios~\cite{khan2017cross}: 1. No User-No Item overlap (NU-NI); 2. User–No Item overlap (U-NI); 3. No User–Item overlap (NU-I); and 4. User–Item overlap (U-I).
Among different scenarios, the U-NI scenario is a popular one in CDR where items from different domains have no overlap and the users are shared~\cite{khan2017cross,fernandez2012cross,cantador2015cross}. It has practical use in real world recommender systems such as Amazon or Taobao, where the same user could buy products in different categories (\emph{e.g.} movies, books and CDs) that refer to different domains~\cite{cremonesi2011cross,hu2018conet}.
Research in this scenario has been proved to provide better shopping service for online users~\cite{fernandez2011generic,gao2013cross,cantador2015cross}.

In the U-NI scenario, researchers have studied various perspectives to transfer knowledge across domains and better predict user behaviors. For example, Li et al.~\cite{li2009transfer} introduced a shared cluster-level rating model which defines a rating function for the latent user- and item-cluster variables to transfer knowledge. Others~\cite{singh2008relational,hu2013personalized,mirbakhsh2015improving} utilized different variants of matrix factorization (MF) approaches. However, most clustering and MF-based methods cannot capture the complex pattern in user-item interactions. Thus, some deep learning based methods~\cite{elkahky2015multi,lian2017cccfnet,hu2018conet,yuan2019darec} emerged to improve the knowledge transfer and mine the complex patterns indicated by user-item interactions.
For example, a deep cross connection network is designed in \cite{hu2018conet} to learn and transfer the shared interaction knowledge among domains. DARec~\cite{yuan2019darec} employs the domain adaptation technique in~\cite{ganin2016domain,hong2020domain} to learn domain-invariant user representations for CDR and it has achieved remarkable performance. Further, some works~\cite{man2017cross,ijcai2018,bi2020dcdir} propose to model domain-specific features of user representations by employing a multi-layer perceptron (MLP) as the mapping function across domains.

Recent knowledge transfer works~\cite{dumoulin2016adversarially,du2018multi,chen2019multivariate} indicate that modeling the joint distribution of different domain samples facilitates better knowledge transfer since joint distribution inherently captures the correlation of different domain samples. Similarly, modeling the joint distribution of user behaviors across domains is crucial in CDR, because the user behaviors in different domains are correlated together. Recent works~\cite{singh2008relational,elkahky2015multi,yuan2019darec} based on the idea of shared-user representation attempt to model the above joint distribution.
With the idea of shared-user representation, the CDR model resorts to learn the overlapped features for feature alignment, which usually leads to compromise the domain-specific features that help to better predict user behaviors ~\cite{man2017cross,2018_c7c46d4b}. In this context, the CDR model may not well do the feature alignment because of the user behavior prediction loss, as well as be hard to learn the domain-specific features for improved recommendation performance~\cite{chang2020domain}. 
Several works~\cite{man2017cross,ijcai2018,bi2020dcdir} have proposed to model the domain-specific features in addition to the overlapped features by employing MLP as the mapping function of user representations in each domain. However, the sparse data in user behaviors easily causes over-fitting for the mapping function~\cite{man2017cross}. EMCDR~\cite{man2017cross} proposes to choose users with dense behaviors to learn the MLP, while it requires heuristic human knowledge for choosing users. ATLRec~\cite{li2020atlrec} improves them with intuitively designed network architecture, 
but also requires heuristic ways for network design. 

\begin{figure*}[t]
\centering
\includegraphics[width=10.0cm]{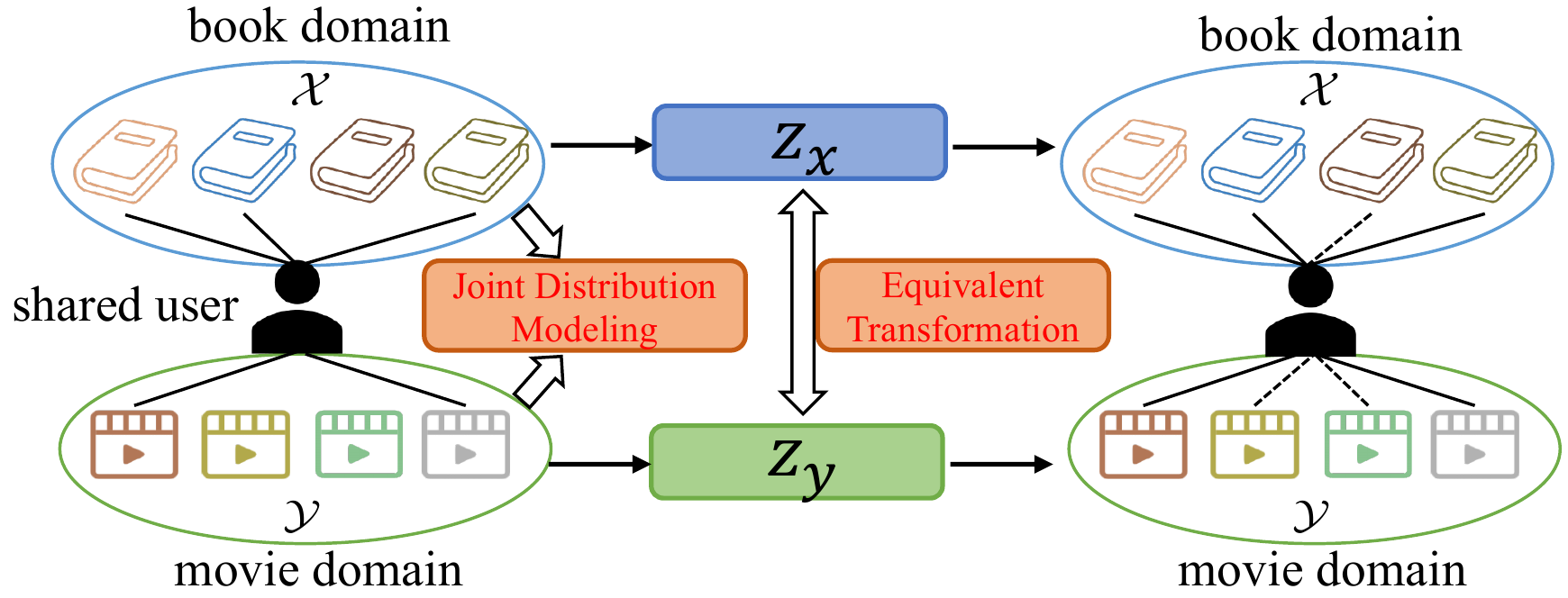}
\caption{Our equivalent transformation based model for recommendation in the U-NI scenario. In this figure, solid lines mean observed user-item interactions and dashed lines are the interactions we aim to predict with probabilities.
Given an user, $z_{x}$ and $z_{y}$ are the preference representations inferred from the user-item interactions in domain $\mathcal{X}$ and domain $\mathcal{Y}$, respectively. The proposed ETL has an equivalent transformation between $z_{x}$ and $z_{y}$, and models the joint distribution of the user behaviors across domains.}
\label{figure:intro_figure}
\end{figure*}

In this paper, we attempt to learn both the overlapped and domain-specific features for CDR in a more principled way. In particular, we assume that each user's preferences of different domains can be mutually converted to each other with equivalent transformation. Then, we propose an equivalent transformation learner (ETL) which models the joint distribution of user behaviors across domains.
The equivalent transformation in ETL relaxes the idea of shared-user representation and enables a user's preferences across domains to have the capacity of
preserving the domain-specific features as well as learning the overlapped features.
Moreover, the proposed ETL has no requirement of carefully selecting training samples like~\cite{man2017cross,ijcai2018}.
Figure~\ref{figure:intro_figure} shows our idea of equivalent transformation based recommendation method. We show that when using ETL, the recommendation accuracy is largely improved compared to state-of-the-art methods on several public benchmarks. The contributions are summarized as follows:
\begin{itemize}
\setlength{\itemsep}{0pt}
\setlength{\parsep}{0pt}
\setlength{\parskip}{0pt}
 \item We highlight the importance of modeling the joint distribution of user behaviors across domains for CDR;
 \item We make an equivalent transformation assumption and further propose a novel method named ETL that models both the overlapped and domain-specific features in a joint distribution matching scheme. The proposed model works in a more principled way and does not require choosing training users or intuitively designing the networks;
 \item Extensive experiments on three public benchmarks demonstrate the effectiveness of the proposed ETL. Empirically, the results of designed experiments show that ETL has better ability to capture the overlapped and domain-specific features of user preferences in CDR.
\end{itemize}

The rest of this paper is organized as follows. Section 2 introduces the preliminary. Section 3 gives the problem definition and the demonstration of the proposed method. Section 4 provides the experiments and analysis to verify the effectiveness of the proposed model. Finally, conclusion and future work are given in Section 5.
\section{Background and Related Work}
Recommender systems has been an important infrastructure for web services~\cite{chen2019cascading,xu2019decoupled,cui2018variational,yao2021device}, among which, CDR emerged as one technique to relieve the sparsity problem on recommendation~\cite{cantador2015cross}.
CDR has been investigated from both the non-deep-learning aspect and deep-learning aspect. In addition, deep generative modeling is also introduced here to better demonstrate the proposed method in Section~3.
\subsection{Non-deep-learning Based CDR}
Early CDR methods mainly concentrate on neighborhood-based solutions~\cite{berkovsky2007cross}. However, the neighborhood-based CDR methods tend to detect localized relationships and fail to capture the totality of weak signals implied by the user's whole interactions~\cite{hu2013personalized}. Consequently, these methods are dominated by MF-based~\cite{singh2008relational,hu2013personalized,lian2017cccfnet} and clustering-based~\cite{mirbakhsh2015improving,gao2013cross} methods. For example, collective matrix factorization (CMF)~\cite{singh2008relational} factorizes multiple user-item interaction matrices from different domains by sharing the user latent factor. CCCFNet~\cite{lian2017cccfnet} combines collaborative filtering and content-based filtering into one unified matrix factorization framework. Considering clustering is one practical technique to alleviate the sparsity problem in single domain recommendation~\cite{linden2003amazon,mirbakhsh2013clustering}, cluster-level matrix factorization~\cite{mirbakhsh2015improving} leverages K-means to capture the shared patterns between the cluster of users and the cluster of items from different domains. 

Although the neighborhood-based, MF-based and clustering-based methods have achieved promising results, they are limited for CDR due to the following reasons. First, most models are linear, which fail to extract the complex patterns in user-item interactions. Second, the relatively dense information of other domains is required to augment the target domains~\cite{yuan2019darec}. In contrast, the proposed ETL is a deep learning model that can extract the complex patterns and has no requirement of a relatively dense information for other domains.

\subsection{Deep-learning Based CDR}
Owing to the superior representation learning ability of deep learning, many deep-learning based CDR methods~\cite{elkahky2015multi,lian2017cccfnet,hu2018conet,kanagawa2019cross,yuan2019darec} have been proposed. Deep learning in CDR not only explores how to capture the complex patterns in user-item interactions, but also pursues a more effective way to transfer knowledge across domains~\cite{pan2009survey}. 
Inspired by the concept of modeling data in semantic space of Deep Structured Semantic models (DSSM)~\cite{huang2013learning}, Elkahky et al.~\cite{elkahky2015multi} proposed a deep learning approach to project users and items in a shared semantic space, and recommend items that have maximum similarity with users in that space. 
CoNet~\cite{hu2018conet} employs deep cross connection network to transfer knowledge between user-item interactions from different domains.
PPGN~\cite{zhao2019cross} propagates user preferences with graph neural networks in CDR. Motivated by dual learning~\cite{xia2016dual}, Li et al.~\cite{li2019DDTCDR} introduced a deep dual transfer network named DDTCDR to enhance the bidirectional knowledge in CDR. Moreover, an orthogonal mapping is used in DDTCDR to extract user preferences across domains while preserving relations between users in the latent space. It is worthwhile to point that the orthogonal mapping is a special case of the equivalent transformation in ETL, and we also study other kinds of equivalent transformation in the experiments of Section~\ref{sec:diff_trans}.
Inspired by domain adaptation in~\cite{ganin2016domain}, several works~\cite{kanagawa2019cross,yuan2019darec} employ domain adaptation techniques to perform knowledge transfer and learn the overlapped features.
Further, some researchers point out that the domain-specific features of user preferences are also important in CDR. They employ MLP as the mapping function across domains to offer learning flexibility for user representations in each domain~\cite{man2017cross,ijcai2018,bi2020dcdir}. However, although MLP increases the learning flexibility, it is easier to have the over-fitting problem for the MLP due to the data sparisity in recommendation~\cite{man2017cross}. Correspondingly, EMCDR~\cite{man2017cross} proposes to choose users with dense behaviors for learning the MLP. 
DCDCSR~\cite{ijcai2018} defines a metric to measure the sparse degree and incorporates the sparse degree when learning the representation mapping across domains. ATLRec~\cite{li2020atlrec} employs different MLP functions to learn the domain-shareable and domain-specific features by an adversarial transfer learning based scheme. 
DCDIR~\cite{bi2020dcdir} utilizes additional knowledge graphs to learn sparsity-insensitive representations for the mapping. Other researchers are also motivated by incorporating side information (\emph{e.g.} item review and user profile) in CDR~\cite{gunasekar2012survey,hu2018mtnet,fu2019deeply}. For example, Hu et al.~\cite{hu2018mtnet} recommended to combine the rich text information (\emph{e.g.} review and title) together to augment CDR. Further, the specific CDR applications include cross platform social e-commerce~\cite{lin2019cross,10.1145/3017429}, multi-modal video recommendation~\cite{10.1145/1961209.1961213} and cross domain collaboration recommendation~\cite{tang2012cross}.

In summary, previous methods~\cite{singh2008relational,lian2017cccfnet,ganin2016domain,yuan2019darec} are mainly based on the idea of shared-user representation, and focus on learning the overlapped features. In this paper, the proposed ETL advocates to learn both the overlapped and domain-specific features for CDR. Compared to recent works that learns the domain-specific features by an MLP mapping function that relies on selected user behaviors~\cite{man2017cross},
the proposed ETL does not require careful choices of training samples. When considering the CDR works based on side information (\emph{e.g.} item review, user profile), our method focuses on better knowledge transfer by employing user-item interactions. 
\subsection{Deep Generative Modeling}
Deep generative modeling is a powerful technique for modeling data distributions. It has been considerably discussed by various methodologies and extended to many applications. In recent years, variational inference~\cite{hoffman2013stochastic} and generative adversarial learning~\cite{goodfellow2014generative} are two representative techniques for deep generative modeling. Variational inference imposes distribution matching in a probabilistic way with specified objective functions. For example, VAE~\cite{kingma2013auto} introduces Kullback-Leibler (KL) divergence to match the latent posterior distribution with the Gaussian prior. Based on VAE, various variants~\cite{zhao2017infovae} are proposed to improve the encoding and reconstruction performance, together with better distribution modeling manners. 
For example, Zhao et al.~\cite{zhao2017infovae} proposed InfoVAE with mutual information maximization in the objective function. 
InfoVAE improves the quality of the variational posterior and encourage better distribution modeling.
GAN~\cite{goodfellow2014generative,nowozin2016f,mao2017least,arjovsky2017wasserstein} is another hot deep generative modeling technique. GAN contains a generator and a discriminator, where the discriminator tries to distinguish the real samples with the fake samples and the generator tries to confuse the discriminator. Adversarial learning in GAN has the advantage of measuring the distribution distance in a more elegant way by binary classification and frees researchers from the painful practice of defining a tricky objective function. Several works~\cite{goodfellow2014generative,nowozin2016f,chen2016infogan} have pointed that the adversarial loss in GAN actually minimizes the Jensen-Shannon divergence (JSD) between the data distribution and the generator distribution. Original GAN has risk facing the vanishing gradient and mode collapse problem. To solve this, some works~\cite{mescheder2017adversarial,srivastava2017veegan,liu2017unsupervised,huang2018munit} 
combine VAE and GAN theories together for better distribution matching. For example, UNIT~\cite{liu2017unsupervised} and MUNIT~\cite{huang2018munit} are proposed to do image-to-image translation task. 

These VAE+GAN works are based on a shared-latent space assumption to do the distribution matching. However, in CDR, the user behaviors across domains are correlated but do not fully overlapped, and the domain-specific features are important for recommendation. In ETL, an equivalent transformation based distribution matching scheme is proposed. 
\begin{table}[]
\caption{Notations in this paper.}
\label{table:notations}
\begin{tabular}{ll}
\hline
Notation & Description \\ \hline
$\mathcal{U}$        & a set of users in CDR            \\
$\mathcal{X}$ and $\mathcal{Y}$       & two different domains             \\
$I^\mathcal{X}$      & the item set in $\mathcal{X}$ domain             \\
$I^\mathcal{Y}$      & the item set in $\mathcal{Y}$ domain             \\
$N$             & the number of users   \\
$M$            & the number of items in $\mathcal{X}$ domain  \\
$T$            & the number of items in $\mathcal{Y}$ domain  \\
$R^\mathcal{X}$                  & the user behavior matrix in $\mathcal{X}$ domain           \\ 
$R^\mathcal{Y}$                  & the user behavior matrix in $\mathcal{Y}$ domain           \\ 
$x_{i}$    & the behavior of user $i$ in $\mathcal{X}$ domain \\
$y_{i}$    & the behavior of user $i$ in $\mathcal{Y}$ domain \\
$E_{x},E_{y}$    & the encoders of user behaviors in different domains \\
$D_{x},D_{y}$    & the decoders of user latent codes in different domains \\
$z_{x}$    & the latent codes of a user in $\mathcal{X}$ domain \\
$z_{y}$    & the latent codes of a user in $\mathcal{Y}$ domain \\
$W_{x}$, $W_{y}$ and $W$       & trainable matrices to approximate the equivalent transformation \\
$\mathcal{D}_{x},\mathcal{D}_{y}$    & the discriminators in adversarial distribution matching \\
$d$        & the latent embedding dimension \\ 
$\lambda$    & the hyper-parameter of equivalent transformation constraint  \\
$\eta$    &  the hyper-parameter on $\mathcal{L}_{PRL}$  \\
$\phi_{x},\phi_{y}$    &  parameters of the encoders  \\
$\theta_{x},\theta_{y}$    &  parameters of the decoders  \\
$\psi_{x},\psi_{y}$    &  parameters of the discriminators  \\
$\mathcal{L}_{JRL}$    & the joint reconstruction loss \\
$\mathcal{L}_{PRL}$    & the prior regularization loss \\
$\mathcal{L}_{ETL}$    & the objective function of ETL  \\
\hline
\end{tabular}
\end{table}

\section{Method}
\begin{figure*}[t]
\centering
\includegraphics[width=13.5cm]{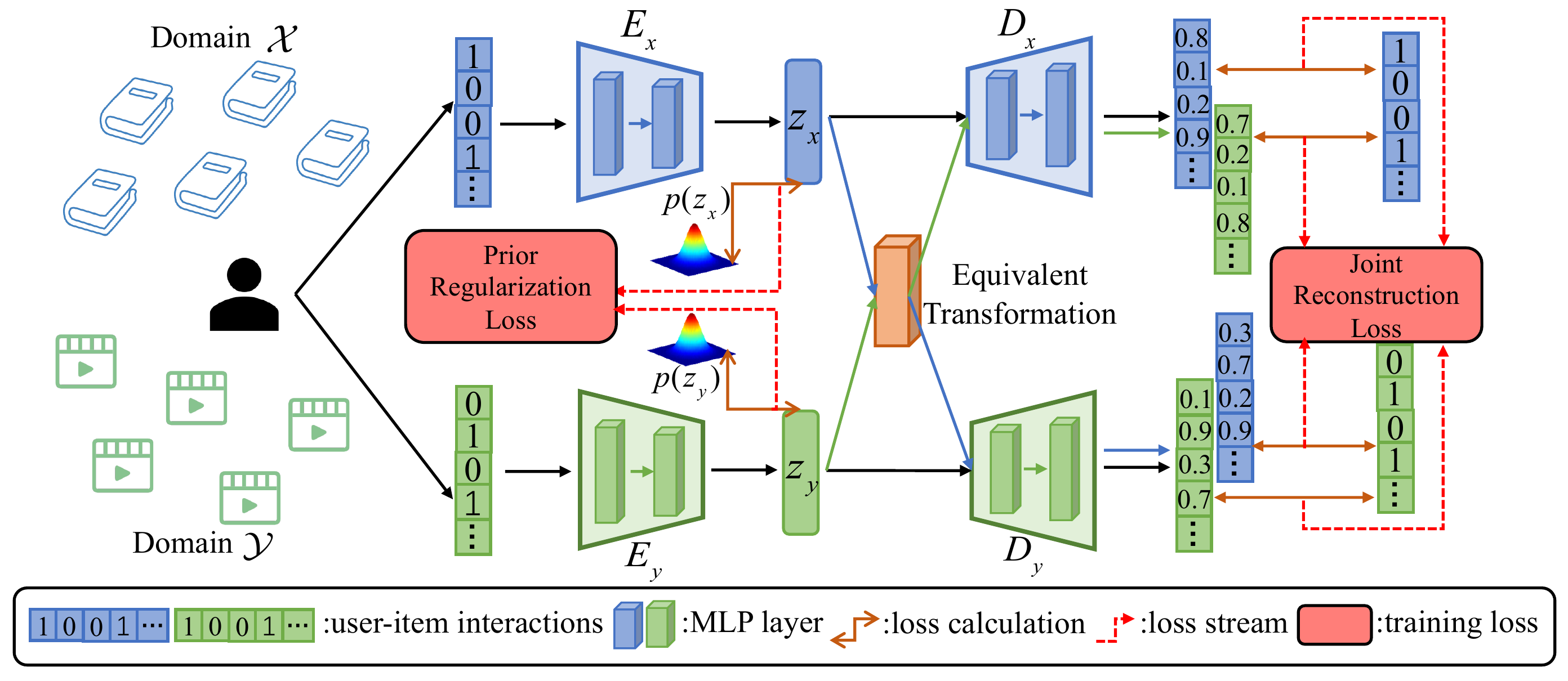}
\caption{The architecture of ETL. ETL encodes user behaviors in two domains with different encoders and then decodes the latent codes to user behaviors in each domain. The joint reconstruction loss and a prior regularization loss facilitates knowledge transfer between two domains and benefits the user behavior prediction.}
\label{figure:model_architecture}
\end{figure*}

In this section, we first give the problem definition. Then, details about the proposed method is introduced. The model architecture of the proposed method is shown in Figure~\ref{figure:model_architecture}.
\subsection{Problem Definition}
In this paper, we focus on the U-NI scenario~\cite{khan2017cross} in CDR. In this setting, different domains $\mathcal{X}$ and $\mathcal{Y}$ have the same set of users $\mathcal{U}=\{U_{1},U_{2},...,U_{N}\}$ where $N$ denotes the number of users. The item set of domain $\mathcal{X}$ and $\mathcal{Y}$ respectively is $I^{\mathcal{X}}=\{I^{\mathcal{X}}_{1},I^{\mathcal{X}}_{2},...,I^{\mathcal{X}}_{M}\}$ and $I^{\mathcal{Y}}=\{I^{\mathcal{Y}}_{1},I^{\mathcal{Y}}_{2},...,I^{\mathcal{Y}}_{T}\}$, where $M$ and $T$ indicate the number of items in domain $\mathcal{X}$ and $\mathcal{Y}$, respectively. 
The user-item interactions of domain $\mathcal{X}$ could be represented by a matrix $R^{\mathcal{X}}\in \mathbb{R}^{N\times M}$ where the values are explicit feedback, such as ratings, or implicit feedback, such as clicks. Similarly, the user-item interactions of domain $\mathcal{Y}$ is indicated by $R^{\mathcal{Y}}\in \mathbb{R}^{N\times T}$. 
Usually, $R^{\mathcal{X}}$ and $R^{\mathcal{Y}}$ are very sparse since a user only interacts with a small subset of items in each domain. 
The goal of CDR is to improve the recommendation accuracy for users in domain $\mathcal{X}$ and $\mathcal{Y}$. 
Unlike~\cite{yuan2019darec}, we do not distinguish a source domain or a target domain since the recommendation task for each domain is performed in an unified method here. The main notations in this paper are summarized in Table~\ref{table:notations}.

\subsection{Overview}
Modeling the joint distribution of user behaviors across domains is essential since the behaviors exhibit correlations in CDR. 
By modeling the joint distribution, we learn more representative user preferences which can help to predict the missing interactions.
The proposed ETL is based on modeling the above joint distribution. Thus, we start with the joint distribution to introduce ETL.
We model the joint distribution via maximizing the joint log-likelihood of the observations.

Assume $x_{i}$ and $y_{i}$ are the behaviors for user $U_{i}$ in domain $\mathcal{X}$ and $\mathcal{Y}$ respectively. In other words, $x_{i}$ and $y_{i}$ are the row vectors of $R^{\mathcal{X}}$ and $R^{\mathcal{Y}}$. Let $(x_{i},y_{i})$ be one paired sample for user $U_{i}$. The joint log-likelihood of observations is composed of a sum over the likelihoods of individual data points $\sum_{i=1}^{N}\log p_{\theta}(x_{i}, y_{i})$, where $p_{\theta}(x_{i}, y_{i})$ denotes the probability density function.
\begin{equation}
\label{eq:joint_log_likelihood}
    \log p_{\theta}((x_{1},y_{1}),(x_{2},y_{2}),...,(x_{N},y_{N}))=\sum_{i=1}^{N}\log p_{\theta}(x_{i}, y_{i})
\end{equation}
Borrowing the idea of maximizing the marginal log-likelihood in VAE~\cite{kingma2013auto}, if $z_{x}$ and $z_{y}$ are the latent factors of $x_{i}$ and $y_{i}$ respectively, $\log p_{\theta}(x_{i},y_{i})$ can be formulated as:
\begin{equation}
\label{eq:original_objective}
\begin{split}
    \log p_{\theta}(x_{i},y_{i})=& D_{KL}[q_{\phi}(z_{x},z_{y}|x_{i}, y_{i})||p(z_{x},z_{y}|x_{i}, y_{i})]\\ 
    & +\mathcal{L}(\theta,\phi;x_{i},y_{i})
\end{split}
\end{equation}
where the first term is the KL divergence of the approximate posterior $q_{\phi}(z_{x},z_{y}|x_{i},y_{i})$ from the true posterior $p(z_{x},z_{y}|x_{i},y_{i})$. Since this KL term is non-negative, the second term is the \textit{evidence lower bound} (\textit{ELBO}) on the log-likelihood $\log p_{\theta}(x_{i},y_{i})$. Following the derivation in VAE~\cite{kingma2013auto}, $\mathcal{L}(\theta,\phi;x_{i},y_{i})$ can be written as:
\begin{align}
\label{eq:ELBO}
\mathcal{L}(\theta,\phi;x_{i},y_{i})=&\mathbb{E}_{q_{\phi}(z_{x},z_{y}|x_{i},y_{i})}[\log p_{\theta}(x_{i},y_{i}|z_{x},z_{y})] \nonumber \\
&-D_{KL}[q_{\phi}(z_{x},z_{y}|x_{i},y_{i})||p(z_{x},z_{y})]
\end{align}
where $p_{\theta}(x_{i},y_{i}|z_{x},z_{y})$ denotes the conditional distribution parameterized by $\theta$. The first term in Eq.~\ref{eq:ELBO} indicates the joint reconstruction loss where $z_{x},z_{y}$ encoded from $x_{i},y_{i}$ are used to reconstruct $x_{i},y_{i}$.
The second term in Eq.~\ref{eq:ELBO} indicates the prior regularization loss where $q_{\phi}(z_{x},z_{y}|x_{i},y_{i})$ is expected to match the prior distribution $p(z_{x},z_{y})$. ETL implements the the joint reconstruction loss and the prior regularization loss via a dual auto-encoder structure and an adversarial learning scheme. The architecture of ETL is shown in Figure~\ref{figure:model_architecture}.
In the following sections, we provide details on the two losses, followed by the objective function and implementation. 

\subsection{Joint Reconstruction Loss}
The first term $\mathbb{E}_{q_{\phi}(z_{x},z_{y}|x_{i},y_{i})}[\log p_{\theta}(x_{i},y_{i}|z_{x},z_{y})]$ in Eq.~\ref{eq:ELBO} consists of an approximate posterior $q_{\phi}(z_{x},z_{y}|x_{i},y_{i})$ parameterized by $\phi$ and a conditional distribution $p_{\theta}(x_{i},y_{i}|z_{x},z_{y})$ parameterized by $\theta$. To ease the solution of the posterior $q_{\phi}(z_{x},z_{y}|x_{i},y_{i})$, we employ the mean filed theory~\cite{zhang2018advances} to define the approximate posterior by following recent works~\cite{huang2018munit,zhu2017unpaired,liu2017unsupervised}:
\begin{align}
\label{eq:rewritten_q_phi}
    q_{\phi}(z_{x},z_{y}|x_{i},y_{i})
    \xlongequal[]{\text{def}} q_{\phi_{x}}(z_{x}|x_{i})q_{\phi_{y}}(z_{y}|y_{i})
\end{align}
which means the latent codes $z_{z}$ and $z_{y}$ are encoded from $x_{i}$ and $y_{i}$, respectively.


\textbf{\emph{Equivalent Transformation Assumption}}: 
In order to solve the conditional distribution\\
$p_{\theta}(x_{i},y_{i}|z_{x},z_{y})$, we make the following equivalent transformation (ET) assumption:
\begin{Assumption}
\label{assumption:1}
\textit{In CDR, each user's preferences of different domains are correlated and can be mutually converted to each other with equivalent transformation.}
\end{Assumption}
According to the definition in mathematics~\cite{gustafson1979matrix}, the equivalent transformation between $z_{x}$ and $z_{y}$ is defined as $z_{x}=Q^{-1}z_{y}P$, where $Q,P$ are two invertible matrices. Note that we set $Q$ as an identity matrix $I$ here for simplicity. Thus, if we denote $z_{y\rightarrow x}$ as the equivalently transformed $z_{y}$ and $z_{x\rightarrow y}$ as the equivalently transformed $z_{x}$, we have $z_{y\rightarrow x}=z_{y}W_{x}$, $z_{x\rightarrow y}=z_{x}W_{y}$ and $W_{x}=P,W_{x}W_{y}=I$.
The equivalent transformation assumption defines a mathematical formulation between the user preference in two related domains. It connects the preference representations across domains in a softer way than the idea of shared-user representations (See more in Section~\ref{sec:disscussion_ET}), and thus encourages the model to learn the domain-specific features for each single domain.
Further, according to the graphical model of ETL in Figure~\ref{figure:graphical_model}, we have the following proposition:
\begin{figure*}[t]
\centering
\includegraphics[width=5.0cm]{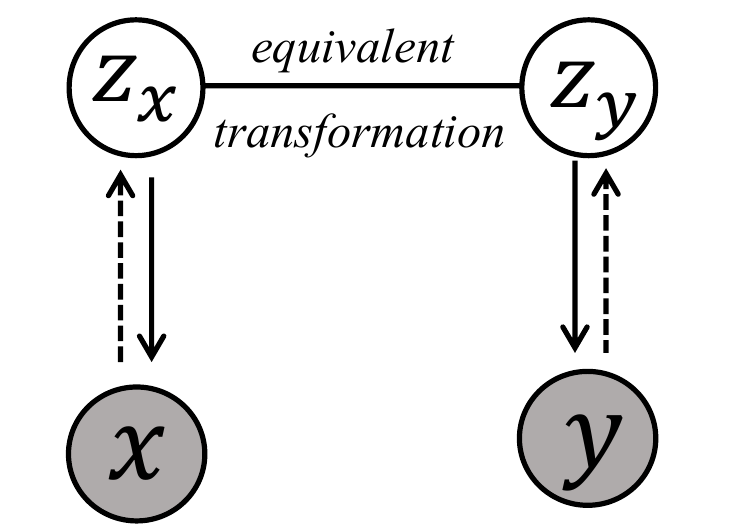}
\caption{The graphical model of our the equivalent transformation based model. In this figure, $z_{x}$ and $z_{y}$ are the latent variables of user representations inferred from the behaviors $x$ of domain $\mathcal{X}$ and $y$ of domain $\mathcal{Y}$, respectively. 
Solid arrows denote the generative process and dashed arrows denote the inference process.}
\label{figure:graphical_model}
\end{figure*}
\begin{Proposition}
\vspace{-3pt}
\label{proposition:1}
\textit{\ding{172}: Given latent variables $z_{x}$, $z_{y}$, the observations $x_{i}$ and $y_{i}$ are conditional independent. \ding{173}: Given the latent variable $z_{x}$, the observation $x_{i}$ and the latent variable $z_{y}$ are conditional independent. \ding{174}: Given the latent variable $z_{y}$, the observation $y_{i}$ and the latent variable $z_{x}$ are conditional independent.}
\vspace{-3pt}
\end{Proposition}
With this proposition and ET assumption, we rewrite the conditional distribution as:
\begin{align}
\label{eq:rewritten_original_conditional}
   p_{\theta}(x_{i},y_{i}|&z_{x},z_{y})\overset{\text{\ding{172}}}{=}p_{\theta_{x}}(x_{i}|z_{x},z_{y})p_{\theta_{y}}(y_{i}|z_{x},z_{y}) \nonumber \\
   &\xlongequal[]{\text{\ding{173}~and~\ding{174}}}p_{\theta_{x}}(x_{i}|z_{x})p_{\theta_{y}}(y_{i}|z_{y}) \nonumber \\
   &=\sqrt{p_{\theta_{x}}(x_{i}|z_{x})p_{\theta_{x}}(x_{i}|z_{x})}\sqrt{p_{\theta_{y}}(y_{i}|z_{y})p_{\theta_{y}}(y_{i}|z_{y})} \nonumber \\
    &\xlongequal[]{\text{ET assumption}}\sqrt{p_{\theta_{x}}(x_{i}|z_{x})p_{\theta_{x}}(x_{i}|z_{y}W_{x})}\sqrt{p_{\theta_{y}}(y_{i}|z_{y})p_{\theta_{y}}(y_{i}|z_{x}W_{y})}
\end{align}

where $\theta_{x}$ is the parameter of a decoder $D_{x}$ and $\theta_{y}$ is the parameter of a decoder $D_{y}$.
Eq.~\ref{eq:rewritten_original_conditional} indicates a dual auto-encoder structure where $z_{x}$ and $z_{y}$ are used to reconstruct $x_{i}$ and $y_{i}$, respectively. Meanwhile, with the equivalent transformation, $z_{x}$ and $z_{y}$ are able to generate $y_{i}$ and $x_{i}$, respectively. $p_{\theta_{x}}(x_{i}|z_{y}W_{x})$ and $p_{\theta_{y}}(y_{i}|z_{x}W_{y})$ indicate the cross domain generation that can help to learn the equivalent transformation. In the end, the equivalent transformation allows $z_{x}$ and $z_{y}$ to have the capacity of maintaining the domain-specific features as well as learning the overlapped features by this dual auto-encoder structure.

\textbf{\emph{Specification of Equivalent Transformation}}:
Different transformations may have different impacts on knowledge transfer, and they are discussed with experiments in Section~\ref{sec:diff_trans}. In ETL, inspired by~\cite{gao2013cross}, we consider that the equivalent transformation in CDR should avoid false correlations between users. Thus, the orthogonal transformation is employed here since it preserves the inner product of vectors, namely it keeps the user similarities across domains. 
According to the definition of orthogonal transformation, $W_{x}$ and $W_{y}$ satisfies $W_{x}=W$ and $W_{y}=W^{T}$, where $W\in \mathbb{R}^{d\times d}$ is the trainable orthogonal mapping matrix. 

Taking the above into summary, we rewrite $\mathbb{E}_{q_{\phi}(z_{x},z_{y}|x_{i},y_{i})}[\log p_{\theta}(x_{i},y_{i}|z_{x},z_{y})]$ in Eq.~\ref{eq:ELBO} as:
\begin{align}
\label{eq:joint_recon_term}
\mathbb{E}_{q_{\phi}(z_{x},z_{y}|x_{i},y_{i})}[\log p_{\theta}(x_{i},y_{i}|z_{x},&z_{y})]=\frac{1}{2}\big \{\mathbb{E}_{q_{\phi_{x}}(z_{x}|x_{i})}[\log p_{\theta_{x}}(x_{i}|z_{x})]
+\mathbb{E}_{q_{\phi_{y}}(z_{y}|y_{i})}[\log p_{\theta_{x}}(x_{i}|z_{y}W)] \nonumber \\
&+\mathbb{E}_{q_{\phi_{y}}(z_{y}|y_{i})}[\log p_{\theta_{y}}(y_{i}|z_{y})]+\mathbb{E}_{q_{\phi_{x}}(z_{x}|x_{i})}[\log p_{\theta_{y}}(y_{i}|z_{x}W^{T})]\big \}
\end{align}

The optima of Eq.~\ref{eq:joint_recon_term}  is the same as that of Eq.~\ref{eq:joint_recon_term} multiplied by a constant. Thereby, for simplified expression, we write the joint reconstruction loss as:
\begin{align}
\label{eq:objective_JRL}
\min_{\phi_{x},\phi_{y},\theta_{x},\theta_{y},W} \mathcal{L}_{JRL}&=-\mathbb{E}_{q_{\phi_{x}}(z_{x}|x_{i})}[\log p_{\theta_{x}}(x_{i}|z_{x})] \nonumber \\
 &-\mathbb{E}_{q_{\phi_{y}}(z_{y}|y_{i})}[\log p_{\theta_{x}}(x_{i}|z_{y}W)] \nonumber \\
    &-\mathbb{E}_{q_{\phi_{y}}(z_{y}|y_{i})}[\log p_{\theta_{y}}(y_{i}|z_{y})] \nonumber \\
    &-\mathbb{E}_{q_{\phi_{x}}(z_{x}|x_{i})}[\log p_{\theta_{y}}(y_{i}|z_{x}W^{T})] \nonumber \\
    + \lambda &(||z_{x}-z_{x}W^{T}W||_{F}^{1}+||z_{y}-z_{y}WW^{T}||_{F}^{1})
\end{align}
where the last term indicates the regularization loss for equivalent transformation when the transformation is specified as orthogonal mapping. $\lambda$ is the hyper-parameter to weight the importance of the regularization. $\{\phi_{x},\phi_{y}\}$ and $\{\theta_{x},\theta_{y}\}$ are the parameters of encoders and decoders, respectively. 

\subsection{Prior Regularization Loss}
\label{sec:PRL}
The second term $D_{KL}[q_{\phi}(z_{x},z_{y}|x_{i},y_{i})||p(z_{x},z_{y})]$ involves joint prior $p(z_{x},z_{y})$ which indicates a complex prior for $z_{x},z_{y}$. In this paper, we set $p(z_{x},z_{y})=p(z_{x})p(z_{y})$ for simplicity\footnote{The complex prior for this CDR problem could be explored in the future with the guidance of recent works~\cite{tomczak2017vae,rezende2015variational,yin2018semi}.}. 
Taking Eq.~\ref{eq:rewritten_q_phi} into consideration, the prior regularization term is:
\begin{align}
\label{eq:kl_new_form}
 D_{KL}[q_{\phi}(z_{x},z_{y}|x_{i},y_{i})||p(z_{x},z_{y})]&=D_{KL}[q_{\phi_{x}}(z_{x}|x_{i})||p(z_{x})]\nonumber \\
    +D_{KL}&[q_{\phi_{y}}(z_{y}|y_{i})||p(z_{y})]
\end{align}
where $p(z_{x})$ and $p(z_{y})$ are the prior distributions. Eq.~\ref{eq:kl_new_form} states the regularization that matches $q_{\phi_{x}}(z_{x}|x_{i})$ to prior $p(z_{x})$ and matches $q_{\phi_{y}}(z_{y}|y_{i})$ to prior $p(z_{y})$.

Since it is not easy to derive explicit formulations for some complex priors in KL-divergence, ETL employs the adversarial distribution matching that can impose an arbitrary prior distribution for the latent codes without hard derivation~\cite{makhzani2015adversarial}. Namely, we utilize adversarial learning to perform the distribution matching between $q_{\phi_{x}}(z_{x}|x_{i})$ (\textit{resp.} $q_{\phi_{y}}(z_{y}|y_{i})$) and $p(z_{x})$ (\textit{resp.} $p(z_{y})$).
Following~\cite{makhzani2015adversarial}, the prior regularization loss
can be formulated with the following adversarial learning scheme:
\begin{align}
\label{eq:objective_PRL}
  \min_{\psi} \max_{\phi} \mathcal{L}_{PRL}=&-\mathbb{E}_{z_{x}\sim p(z_{x})}[\log \mathcal{D}_{x}(z_{x})] \nonumber\\
  &-\mathbb{E}_{z_{x}\sim q_{\phi_{x}(z_{x}|x_{i})}}[\log (1-\mathcal{D}_{x}(z_{x}))] \nonumber \\
  &-\mathbb{E}_{z_{y}\sim p(z_{y})}[\log \mathcal{D}_{y}(z_{y})] \nonumber\\
  &-\mathbb{E}_{z_{y}\sim q_{\phi_{y}(z_{y}|y_{i})}}[\log (1-\mathcal{D}_{y}(z_{y}))]
\end{align}
where $\phi=\{\phi_{x},\phi_{y}\}$ shares the same definition in Eq.~\ref{eq:objective_JRL}
and $\psi=\{\psi_{x},\psi_{y}\}$ are the parameters of the discriminators $\mathcal{D}_{x},\mathcal{D}_{y}$. $p(z_{x})$ and $p(z_{y})$ are the prior distributions of $z_{x}$ and $z_{y}$ respectively. 

The employed adversarial distribution matching in Eq.~\ref{eq:objective_PRL} has several advantages compared to the KL-divergence in Eq.~\ref{eq:kl_new_form}. The KL-divergence tries to match $q_{\phi_{x}}(z_{x}|x_{i})$ to prior $p(z)$, which will have risk to lose the information from input $x_{i}$. By contrast,
the adversarial distribution matching in latent space makes the posterior $q_{\phi_{x}}(z_{x}|x_{i})$ to be the aggregated posterior $q_{\phi_{x}}(z_{x})$, which encourages $z_{x}$ to match the whole distribution of $p(z_{x})$~\cite{makhzani2015adversarial,makhzani2018implicit}. Accordingly, $z_{y}$ can match the whole distribution of $p(z_{y})$ in similar way.
Meanwhile, the mode collapse problem in adversarial learning could be avoided since our method involves a reconstruction loss which encourages the latent embeddings to match both the prior and the entire true data distribution~\cite{srivastava2017veegan}.

\subsection{Objective Function and Implementation}
\label{sec:objective_implementation}
Taking Eq.~\ref{eq:objective_JRL} and Eq.~\ref{eq:objective_PRL} into account, maximizing the \textit{ELBO} in Eq.~\ref{eq:ELBO} can be performed via optimizing the following objective function:
\begin{equation}
\label{eq:objective_ETL}
    \min_{\Theta} \max_{\Phi} \mathcal{L}_{ETL}=\mathcal{L}_{JRL}+\eta \mathcal{L}_{PRL}
\end{equation}
where $\Theta=\{\phi_{x},\phi_{y},\theta_{x},\theta_{y},\psi_{x},\psi_{y},W\}$ and $\Phi=\{\phi_{x},\phi_{y}\}$ are the network parameters. $\eta$ is the hyper-parameter to weight the importance of the prior regularization term.

The architecture of ETL is shown in Fig.~\ref{figure:model_architecture}, where $E_{x}$ is the encoder for $q_{\phi_{x}}(z_{x}|x_{i})$ and $E_{y}$ is the encoder for $q_{\phi_{y}}(z_{y}|y_{i})$. Similarly, $D_{x}$ and $D_{y}$ are the decoders for $p_{\theta_{x}}(x_{i}|\cdot)$ and $p_{\theta_{y}}(y_{i}|\cdot)$, respectively. The discriminators $\mathcal{D}_{x}$ and $\mathcal{D}_{y}$ are designed for adversarial learning. In our implementation, all $E_{x},E_{y},D_{x},D_{y},\mathcal{D}_{x},\mathcal{D}_{y}$ are two-layer MLP with \textit{Relu} as the non-linear activation function. We use standard Gaussian distributions such that $p(z_{x})\sim \mathcal{N}(0,1)$ and $p(z_{y})\sim \mathcal{N}(0,1)$, which is a common choice in recent adversarial learning based methods~\cite{goodfellow2014generative,makhzani2015adversarial,srivastava2017veegan}. It is also worthwhile to point out that although both $p(z_{x})$ and $p(z_{y})$ follow the standard Gaussian distribution, it does not mean $z_{x}$ and $z_{y}$ are in the same latent space and it will not break our motivation of modeling both features of user preferences in CDR. Moreover, the reconstruction loss between the predictions and true data could be \textit{MSE} if the user-item interactions are explicit feedback and \textit{binary cross entropy} if the user-item interactions are implicit feedback.

\subsection{Discussion of Equivalent Transformation}
\label{sec:disscussion_ET}
In this section, we have some discussion about the equivalent transformation to provide deep insight on how ETL connects to and differs from previous methods. 
Although recent methods~\cite{singh2008relational,lian2017cccfnet,elkahky2015multi,kanagawa2019cross,yuan2019darec} with the idea of shared-user representation do not mention this, they attempt to model the joint distribution of user behaviors across domains. According to the coupling theory~\cite{lindvall2002lectures}, if we want to model a joint distribution of two marginal observations $\mathcal{X},\mathcal{Y}$, an assumption is indispensable to describe the relationship of the corresponding latent factors $z_{x}$ and $z_{y}$.
The assumption can be expressed as $z_{x}=z_{y}W_{x},z_{y}=z_{x}W_{y}$ (or as non-linear formulation). There are two possible formulations which are \emph{unconstrained} and \emph{constrained}. The \emph{unconstrained} formulation means there is no constraint between $W_{x}$ and $W_{y}$, and it is similar to the idea of using MLP as the mapping function across domains~\cite{man2017cross,bi2020dcdir}. While the \emph{constrained} one imposes constraints on $W_{x}$ and $W_{y}$. 

In CDR, $z_{x}$ and $z_{y}$ in different domains have correlations. Thus, in ETL, we adopt the \emph{constrained} formulation which specifically is our ET assumption. This assumption involves the equivalent transformation with $W_{x}W_{y}=I$, which degenerates to the idea of shared-user representation in~\cite{singh2008relational,elkahky2015multi,yuan2019darec} when $W_{x}=W_{y}=I$. However, $W_{x}=W_{y}=I$ would make the domain-specific features be suppressed. In contrary, ETL with ET assumption allows a user's preferences in different domains to have the capacity of learning the domain-specific features as well as the overlapped features. It is worthwhile to point out that the ET assumption ensures cross domain generation that facilitates better knowledge transfer.
Different variants of the \emph{unconstrained} and \emph{constrained} formulations are discussed with experiments in Section~\ref{sec:diff_trans}. 
There may exist two domains where the user behaviors have no relation or nearly no relation so that it is inappropriate to transfer knowledge and apply our equivalent transformation based model. An intuitive way to judge whether the ETL model can work in two domains is to see whether the two domains are different and meanwhile have shared attributes that can connect the user behaviors. For example, Movie and Book domain are two different domains and have obvious overlapped attributes such as theme (\emph{e.g.} love, war). A user likes love movies has high probability of reading romance novels.

\subsection{Time Complexity Analysis}
Stochastic training of DNN methods involves two steps, the forward and backward computations.
ETL supports the mini-batch training and the time cost lies in the joint reconstruction term and the prior regularization term.
We thus decompose the time complexity of ETL into two parts, namely the time complexity of the joint reconstruction term and the prior regularization term. In each batch of ETL, the joint reconstruction term encodes user behaviors into latent codes and then decodes the latent codes into user behaviors. If we denote $B$ as the batch size, $M$ and $T$ as the number of items in each domain, then the complexity of the joint reconstruction term is $\mathcal{O}(B(M+T))$. The prior regularization term imposes prior distribution on the latent codes. If we denote the latent dimension as $d$, the complexity of the prior regularization term is $\mathcal{O}(Bd)$. In summary, the time complexity of ETL is $\mathcal{O}(B(M+T)+Bd)$. 
In addition, user behaviors in each domain usually are extremely sparse vectors. This indicates most values in the $M-$dimensional and $T-$dimensional user behavior vectors are zeros.
Then the sparse vectors can be fast calculated by the sparse matrix multiplication in Pytorch or Tensorflow. If we denote the average number of non-zero values of the $M-$dimensional and $T-$dimensional user behavior vectors in a batch as $c$, we have $c\ll \frac{1}{2}(M+T)$. And the complexity of the joint reconstruction term can be largely reduced to $\mathcal{O}(2Bc)$. In this case, the complexity of ETL is $\mathcal{O}(2Bc+Bd)$, which ensures that ETL can work on much larger datasets. 

Moreover, when applying ETL on multiple domains, a paired combination of each two domains can be used. If we denote the domain number as $N_{d}$, the complexity of ETL is $\mathcal{O}(\frac{N_{d}(N_{d}-1)}{2}(B(M+T)+Bd))$. 
The proposed ETL mainly focuses on how to model the user-item interactions in CDR and we use paired two domains which is a common setting in many works~\cite{hu2018conet,yuan2019darec,li2019DDTCDR} to verify ETL's effectiveness. How to more efficiently learn the model in the multiple-domain case could be explored in the future.

\section{Experiments and Analysis}
\label{sec:experiments}
In this section, we first give the details about the datasets and experimental settings. Then we systemically evaluate ETL via the comparison with recent state-of-the-art methods on multiple public benchmarks. Next, we design two experiments to demonstrate that ETL simultaneously learns the overlapped and domain-specific features in CDR. Finally, we conduct the ablation study on ETL.
\subsection{Dataset Description}
To evaluate the effectiveness of ETL, we utilize three largest benchmarks from Amazon\footnote{http://jmcauley.ucsd.edu/data/amazon/}. The three datasets are Movies and TV (\textbf{Movie}), Books (\textbf{Book}), CDs and Vinyl (\textbf{Music}). Note that these three dataset are benchmarks for CDR and have been used in recent works~\cite{hu2018conet,yuan2019darec}.
Following~\cite{hu2018conet,yuan2019darec}, we make a pairwise combinations amongst the three datasets and find the shared users in each of the two domains for the U-NI CDR scenario~\cite{khan2017cross}. Namely, we obtain \textbf{Movie \& Book}, \textbf{Movie \& Music} and \textbf{Music \& Book} as our experimental datasets. 
Compared to the explicit feedback (\emph{e.g.} the user ratings on items), the implicit feedback (\emph{e.g.} the user clicks or does not click an item) are more common in real-world recommender systems~\cite{hu2008collaborative,oard1998implicit}, we thus focus on the implicit user-item interactions in this paper. 
In other words, the user-item interaction matrices $R^{\mathcal{X}},R^{\mathcal{Y}}$ are binary matrices where the value is $1$ (observed or clicked) if the user interacted with the item and $0$ (unobserved or not clicked) otherwise.
Since the user-item interactions in these benchmarks are ratings ranging from 0 to 5, we convert the ratings of 3,4,5 as positive samples by following~\cite{hu2018conet}. Finally, we filter users and items whose number of interactions is less than 5. The dataset statistics are shown in Table~\ref{table:datasets_statistics}. As shown in Table~\ref{table:datasets_statistics}, both domains in each dataset are extremely sparse with at least 99.86\% interactions are unobserved. It presents a great challenge on most clustering-based and MF-based CDR methods since these methods normally require dense interactions in at least one domain~\cite{yuan2019darec}. In addition, the number of items in \textbf{Movie \& Book} is unbalanced and the density in \textbf{Movie \& Music} is unbalanced, which provides more comprehensive evaluation conditions for different CDR methods.

\begin{table}[t]
\centering
\caption{The statistics of datasets.}
\label{table:datasets_statistics}
\renewcommand{\arraystretch}{1.2}
 \setlength{\tabcolsep}{1.8mm}{ 
  \scalebox{1.0}{
\begin{tabular}{c|c|c|c|c|cc}
\hline
Datasets       & \multicolumn{2}{c|}{\begin{tabular}[c]{@{}c@{}}Movie \& Book\end{tabular}} & \multicolumn{2}{c|}{\begin{tabular}[c]{@{}c@{}}Movie \& Music\end{tabular}} & \multicolumn{2}{c}{\begin{tabular}[c]{@{}c@{}}Music \& Book\end{tabular}} \\ \hline
\#Users        & \multicolumn{2}{c|}{29,476}                                                           & \multicolumn{2}{c|}{15,914}                                                            & \multicolumn{2}{c}{16,267}                                                           \\ \hline
Domain         & Movie                                     & Book                                      & Movie                                      & Music                                     & \multicolumn{1}{c|}{Music}                          & Book                           \\ \hline
\#Items        & 24,091                                    & 41,884                                    & 17,794                                     & 20,058                                    & \multicolumn{1}{c|}{18,467}                         & 23,988                         \\ \hline
\#Interactions & 591,258                                   & 579,131                                   & 416,228                                    & 280,398                                   & \multicolumn{1}{c|}{233,251}                        & 291,325                        \\ \hline
Density       & 0.08\%                                    & 0.05\%                                    & 0.14\%                                     & 0.09\%                                    & \multicolumn{1}{c|}{0.08\%}                         & 0.07\%                         \\ \hline
\end{tabular}
}}
\end{table}
\subsection{Experimental Setup}
\subsubsection{\textbf{Evaluation Protocols}}
In item recommendation, the leave-one-out (LOO) evaluation is widely used~~\cite{he2017neural,hu2018conet,zamani2020learning,du2018online,yuan2019adversarial} and we also use LOO here. In other words, we randomly reserve two items for each user, one as the validation item and the other one as the test item. Following~\cite{he2017neural,hu2018conet}, we randomly sample 99 items that are not interacted by the user as negative items, and then evaluate how the recommender can rank the validation and test item against the negative items. Since we focus on the implicit feedback in recommendation, we adopt three widely used evaluation metrics: hit ratio (HR), normalized discounted cumulative gain (NDCG) and mean reciprocal rank (MRR). The predicted rank list is cut off at $topK=5,10$. A higher value means a better recommendation performance for all three metrics. Also, during training, we save the best trained model according to the 
performance on the validation set and perform testing with the saved model. 
Moreover, all models are run 5 times and the mean value on the test are reported as the model performance. Empirically, we also show the t-test results in Section~\ref{sec:overall_performance} to illustrate that ETL has statistically significant performance over other methods. Results in other sections are also statistically significant and we only report the mean value of 5 runs for simplicity.

\subsubsection{\textbf{Baselines}}
To illustrate the effectiveness, we compare ETL with single domain methods (PMF, CDAE, CFVAE and AAE) and recent cross domain methods (CMF, AAE++, CoNet, sCoNet, ATLRec, DDTCDR and DARec) as follows:
\begin{itemize}
 \item PMF~\cite{mnih2008probabilistic}: Probabilistic matrix factorization is a classic factorization-based method for single domain recommendation, which has been successfully applied in real systems~\cite{koren2009matrix}.
 Since we focus on the implicit feedback here, we replace the original mean square error (MSE) loss in~\cite{koren2009matrix} with binary cross entropy loss for fair comparison.
 \item CDAE~\cite{wu2016collaborative}: Collaborative denoising auto-encoder is a generalization of several auto-encoder based recommendation methods but with more flexible components.
 \item CFVAE~\cite{liang2018variational}: Collaborative variational auto-encoder is a variational auto-encoder model for collaborative filtering.
 \item AAE~\cite{makhzani2015adversarial}: Adversarial auto-encoder combines the recent generative adversarial networks (GAN) and the auto-encoding variational inference. We follow CDAE's setting here for AAE to perform the recommendation task. 
 \item CMF~\cite{singh2008relational}: Collective matrix factorization is a multi-relational learning method that jointly factorizes the user-item interaction matrices of different domains. 
 \item AAE++~\cite{makhzani2015adversarial}: We extend AAE as AAE++ here for CDR. To be specific, AAE++ performs the adversarial auto-encoder for different domains with the same standard Gaussian distribution as prior and different discriminators. It also serves as a variant of our ETL model with no cross generation stream.
 \item CoNet and sCoNet~\cite{hu2018conet}: CoNet transfers knowledge of different domains through a modified cross-stitch neural network. 
 Specifically, it constructs deep cross-connections for the predicted user-item interactions from different domains. sCoNet is a sparse version of CoNet with L1-regularization on the user and item representations.
 \item ATLRec~\cite{li2020atlrec}: ATLRec is an adversarial transfer learning based model that captures domain-shareable and domain-specific features by different neural networks. In particular, it use two domain-specific neural networks to learn the domain-specific features and one shared neural network to capture the domain-shareable features.
 \item DDTCDR~\cite{li2019DDTCDR}: DDTCDR introduces the mechanism of dual learning in CDR and proposes a deep dual transfer network. DDTCDR requires user features and item features with the same dimension as input, and these features are obtained by employing the matrix factorization technique on user-item interaction matrix.
 \item DARec~\cite{yuan2019darec}: DARec introduces domain adaptation techniques~\cite{ganin2014unsupervised} for CDR and
 has achieved remarkable recommendation performance.
 \item ETL-JRL: ETL-JRL is a variant of our ETL model, which only contains the joint reconstruction loss $\mathcal{L}_{JRL}$. 
\end{itemize}

\begin{table*}[]
\centering
\caption{The overall comparison on Movie \& Book. The underlined results are the best performance of baselines. Compared to ETL, the t-test results of other baselines are shown in this table. $\ddagger$ means p-value$<$0.01, $\dagger$ indicates p-value$<$0.05 and $-$ means p-value$>$0.05.}
\label{table:overall_amazon1}
\renewcommand{\arraystretch}{1.2}
 \setlength{\tabcolsep}{0.5mm}{ 
  \scalebox{0.85}{
\begin{tabular}{ccccccccccccc}
\hline
\multicolumn{13}{c}{Movie \& Book}                                                                                                                                                                                                                                                                           \\ \hline
\multicolumn{1}{c|}{topK}      & \multicolumn{6}{c|}{topK=5}                                                                                                                         & \multicolumn{6}{c}{topK=10}                                                                                                    \\ \hline
\multicolumn{1}{c|}{Domain}    & \multicolumn{3}{c|}{Movie}                                               & \multicolumn{3}{c|}{Book}                                                & \multicolumn{3}{c|}{Movie}                                               & \multicolumn{3}{c}{Book}                            \\ \hline
\multicolumn{1}{c|}{Metrics}   & HR              & NDCG            & \multicolumn{1}{c|}{MRR}             & HR              & NDCG            & \multicolumn{1}{c|}{MRR}             & HR              & NDCG            & \multicolumn{1}{c|}{MRR}             & HR              & NDCG            & MRR             \\ \hline
\multicolumn{1}{c|}{PMF}       & 0.4364$^\ddagger$          & 0.3147$^\ddagger$          & \multicolumn{1}{c|}{0.2745$^\ddagger$}          & 0.4003$^\ddagger$          & 0.2961$^\ddagger$          & \multicolumn{1}{c|}{0.2621$^\ddagger$}          & 0.5737$^\ddagger$          & 0.3591$^\ddagger$          & \multicolumn{1}{c|}{0.2928$^\ddagger$}          & 0.5121$^\ddagger$          & 0.3327$^\ddagger$          & 0.2772$^\ddagger$          \\
\multicolumn{1}{c|}{CDAE}      & 0.4660$^\ddagger$          & 0.3471$^\ddagger$          & \multicolumn{1}{c|}{0.3056$^\ddagger$}          & 0.4483$^\ddagger$          & 0.3492$^\ddagger$          & \multicolumn{1}{c|}{0.3157$^\ddagger$}          & 0.5991$^\ddagger$          & 0.3901$^\ddagger$          & \multicolumn{1}{c|}{0.3263$^\ddagger$}          & 0.5640$^\ddagger$           & 0.3851$^\ddagger$          & 0.3315$^\ddagger$          \\
\multicolumn{1}{c|}{CFVAE}      & 0.4587$^\ddagger$          & 0.3396$^\ddagger$          & \multicolumn{1}{c|}{0.3006$^\ddagger$}          & 0.4277$^\ddagger$          & 0.3258$^\ddagger$          & \multicolumn{1}{c|}{0.2918$^\ddagger$}          & 0.5928$^\ddagger$          & 0.3852$^\ddagger$          & \multicolumn{1}{c|}{0.3206$^\ddagger$}          & 0.5508$^\ddagger$           & 0.3646$^\ddagger$          & 0.3091$^\ddagger$          \\
\multicolumn{1}{c|}{AAE}       & 0.4661$^\ddagger$          & 0.3471$^\ddagger$          & \multicolumn{1}{c|}{0.3080$^\ddagger$}          & 0.4457$^\ddagger$          & 0.3509$^\ddagger$          & \multicolumn{1}{c|}{0.3128$^\ddagger$}          & 0.5989$^\ddagger$          & 0.3900$^\ddagger$          & \multicolumn{1}{c|}{0.3269$^\ddagger$}          & 0.5559$^\ddagger$          & 0.3871$^\ddagger$          & 0.3291$^\ddagger$          \\ \hline
\multicolumn{1}{c|}{CMF}       & 0.4433$^\ddagger$          & 0.3224$^\ddagger$          & \multicolumn{1}{c|}{0.2815$^\ddagger$}          & 0.4373$^\ddagger$          & 0.3225$^\ddagger$          & \multicolumn{1}{c|}{0.2848$^\ddagger$}          & 0.5848$^\ddagger$          & 0.3674$^\ddagger$          & \multicolumn{1}{c|}{0.3000$^\ddagger$}          & 0.5583$^\ddagger$          & 0.3616$^\ddagger$          & 0.3009$^\ddagger$          \\
\multicolumn{1}{c|}{AAE++}     & 0.4803$^\ddagger$          & 0.3590$^\ddagger$          & \multicolumn{1}{c|}{0.3189$^\ddagger$}          & 0.4537$^\ddagger$          & \underline{0.3592$^\ddagger$}    & \multicolumn{1}{c|}{\underline{0.3280$^\ddagger$}}    & 0.6098$^\ddagger$          & 0.4009$^\ddagger$          & \multicolumn{1}{c|}{0.3362$^\ddagger$}          & 0.5656$^\ddagger$          & 0.3954$^\ddagger$          & \underline{ 0.3429$^\ddagger$}    \\
\multicolumn{1}{c|}{CoNet}     & 0.3886$^\ddagger$          & 0.2702$^\ddagger$          & \multicolumn{1}{c|}{0.2279$^\ddagger$}    & 0.3451$^\ddagger$          & 0.2316$^\ddagger$          & \multicolumn{1}{c|}{0.2033$^\ddagger$}          & 0.5244$^\ddagger$          & 0.3145$^\ddagger$          & \multicolumn{1}{c|}{0.2464$^\ddagger$}          & 0.4690$^\ddagger$          & 0.2716$^\ddagger$          & 0.2195$^\ddagger$          \\
\multicolumn{1}{c|}{sCoNet}     & 0.3914$^\ddagger$          & 0.2709$^\ddagger$          & \multicolumn{1}{c|}{0.2277$^\ddagger$}    & 0.3408$^\ddagger$          & 0.2322$^\ddagger$          & \multicolumn{1}{c|}{0.2043$^\ddagger$}          & 0.5308$^\ddagger$          & 0.3167$^\ddagger$          & \multicolumn{1}{c|}{0.2463$^\ddagger$}          & 0.4711$^\ddagger$          & 0.2724$^\ddagger$          & 0.2209$^\ddagger$          \\
\multicolumn{1}{c|}{ATLRec}     & 0.3851$^\ddagger$          & 0.2836$^\ddagger$          & \multicolumn{1}{c|}{0.2435$^\ddagger$}    & 0.3452$^\ddagger$          & 0.2587$^\ddagger$          & \multicolumn{1}{c|}{0.2020$^\ddagger$}          & 0.5349$^\ddagger$          & 0.3354$^\ddagger$          & \multicolumn{1}{c|}{0.2607$^\ddagger$}          & 0.4763$^\ddagger$          & 0.3131$^\ddagger$          & 0.2362$^\ddagger$          \\
\multicolumn{1}{c|}{DDTCDR}     & 0.4090$^\ddagger$          & 0.2942$^\ddagger$          & \multicolumn{1}{c|}{ 0.2576$^\ddagger$}    & 0.4008$^\ddagger$          & 0.3153$^\ddagger$          & \multicolumn{1}{c|}{0.2893$^\ddagger$}          & 0.5394$^\ddagger$          & 0.3382$^\ddagger$          & \multicolumn{1}{c|}{0.2732$^\ddagger$}          & 0.5073$^\ddagger$          & 0.3492$^\ddagger$          & 0.3013$^\ddagger$          \\
\multicolumn{1}{c|}{DARec}     & \underline{ 0.4914$^\ddagger$}    & \underline{ 0.3641$^\ddagger$}    & \multicolumn{1}{c|}{\underline{0.3224$^\ddagger$}}          & \underline{ 0.4690$^\ddagger$}    & 0.3591$^\ddagger$          & \multicolumn{1}{c|}{0.3227$^\ddagger$}          & \underline{ 0.6202$^\ddagger$}    & \underline{ 0.4069$^\ddagger$}    & \multicolumn{1}{c|}{\underline{ 0.3401$^\ddagger$}}    & \underline{ 0.5919$^\ddagger$}    & \underline{ 0.3989$^\ddagger$}    & 0.3392$^\ddagger$          \\ \hline
\multicolumn{1}{c|}{ETL-JRL}  & 0.5109$^\ddagger$          & 0.3805$^-$          & \multicolumn{1}{c|}{0.3427$^-$}          & 0.5020$^\ddagger$           & 0.3940$^\dagger$           & \multicolumn{1}{c|}{0.3663$^\ddagger$}          & 0.6412$^-$          & 0.4157$^\ddagger$          & \multicolumn{1}{c|}{0.3600$^-$}            & 0.6266$^\ddagger$          & 0.4221$^\ddagger$          & 0.3819$^\ddagger$          \\
\multicolumn{1}{c|}{ETL}      & \textbf{0.5115} & \textbf{0.3812} & \multicolumn{1}{c|}{\textbf{0.3431}} & \textbf{0.5111} & \textbf{0.3989} & \multicolumn{1}{c|}{\textbf{0.3705}} & \textbf{0.6419} & \textbf{0.4244} & \multicolumn{1}{c|}{\textbf{0.3608}} & \textbf{0.6329} & \textbf{0.4383} & \textbf{0.3861} \\ \hline
\multicolumn{1}{c|}{\%Improv.} & 4.09\%          & 4.69\%          & \multicolumn{1}{c|}{6.42\%}          & 8.97\%          & 11.05\%         & \multicolumn{1}{c|}{12.95\%}         & 3.49\%          & 4.30\%          & \multicolumn{1}{c|}{6.08\%}          & 6.92\%          & 9.87\%          & 12.59\%         \\ \hline
\end{tabular}
}}
\vspace{-5pt}
\end{table*}

\begin{table*}[]
\centering
\caption{The overall comparison on Movie \& Music. The underlined results are the best performance of baselines.
Compared to ETL, the t-test results of other baselines are shown in this table. $\ddagger$ means p-value$<$0.01, $\dagger$ indicates p-value$<$0.05 and $-$ means p-value$>$0.05.}
\label{table:overall_amazon2}
\renewcommand{\arraystretch}{1.2}
 \setlength{\tabcolsep}{0.5mm}{ 
  \scalebox{0.85}{
\begin{tabular}{ccccccccccccc}
\hline
\multicolumn{13}{c}{Movie \& Music}                                                                                                                                                                                                                                                                          \\ \hline
\multicolumn{1}{c|}{topK}      & \multicolumn{6}{c|}{topK=5}                                                                                                                         & \multicolumn{6}{c}{topK=10}                                                                                                    \\ \hline
\multicolumn{1}{c|}{Domain}    & \multicolumn{3}{c|}{Movie}                                               & \multicolumn{3}{c|}{Music}                                               & \multicolumn{3}{c|}{Movie}                                               & \multicolumn{3}{c}{Music}                           \\ \hline
\multicolumn{1}{c|}{Metrics}   & HR              & NDCG            & \multicolumn{1}{c|}{MRR}             & HR              & NDCG            & \multicolumn{1}{c|}{MRR}             & HR              & NDCG            & \multicolumn{1}{c|}{MRR}             & HR              & NDCG            & MRR             \\ \hline
\multicolumn{1}{c|}{PMF}       & 0.4081$^\ddagger$          & 0.2872$^\ddagger$          & \multicolumn{1}{c|}{0.2474$^\ddagger$}          & 0.4505$^\ddagger$          & 0.3350$^\ddagger$          & \multicolumn{1}{c|}{0.2969$^\ddagger$}          & 0.5490$^\ddagger$          & 0.3326$^\ddagger$          & \multicolumn{1}{c|}{0.2261$^\ddagger$}          & 0.5769$^\ddagger$          & 0.3759$^\ddagger$          & 0.3137$^\ddagger$          \\
\multicolumn{1}{c|}{CDAE}      & 0.4191$^\ddagger$          & 0.3093$^\ddagger$          & \multicolumn{1}{c|}{0.2723$^\ddagger$}          & 0.4433$^\ddagger$          & 0.3396$^\ddagger$          & \multicolumn{1}{c|}{0.3053$^\ddagger$}          & 0.5544$^\ddagger$          & 0.3528$^\ddagger$          & \multicolumn{1}{c|}{0.2898$^\ddagger$}          & 0.5662$^\ddagger$          & 0.3792$^\ddagger$          & 0.3225$^\ddagger$          \\
\multicolumn{1}{c|}{CFVAE}      & 0.4318$^\ddagger$          & 0.3110$^\ddagger$          & \multicolumn{1}{c|}{0.2750$^\ddagger$}          & 0.4362$^\ddagger$          & 0.3281$^\ddagger$          & \multicolumn{1}{c|}{0.2884$^\ddagger$}          & 0.5699$^\ddagger$          & 0.3605$^\ddagger$          & \multicolumn{1}{c|}{0.2945$^\ddagger$}          & 0.5646$^\ddagger$          & 0.3663$^\ddagger$          & 0.3082$^\ddagger$          \\
\multicolumn{1}{c|}{AAE}       & 0.4357$^\ddagger$          & 0.3226$^\ddagger$          & \multicolumn{1}{c|}{0.2860$^\ddagger$}           & 0.4557$^\ddagger$          & 0.3445$^\ddagger$          & \multicolumn{1}{c|}{0.3086$^\ddagger$}          & 0.5689$^\ddagger$          & 0.3658$^\ddagger$          & \multicolumn{1}{c|}{0.3023$^\ddagger$}          & 0.5772$^\ddagger$          & 0.3863$^\ddagger$          & 0.3248$^\ddagger$          \\ \hline
\multicolumn{1}{c|}{CMF}       & 0.4309$^\ddagger$          & 0.3025$^\ddagger$          & \multicolumn{1}{c|}{0.2603$^\ddagger$}          & 0.4794$^\ddagger$          & 0.3568$^\ddagger$          & \multicolumn{1}{c|}{0.3166$^\ddagger$}          & 0.5736$^\ddagger$          & 0.3487$^\ddagger$          & \multicolumn{1}{c|}{0.2793$^\ddagger$}          & 0.6124$^\ddagger$          & 0.4011$^\ddagger$          & 0.3349$^\ddagger$          \\
\multicolumn{1}{c|}{AAE++}     & 0.4281$^\ddagger$          & 0.3142$^\ddagger$          & \multicolumn{1}{c|}{0.2754$^\ddagger$}          & 0.4538$^\ddagger$          & 0.3501$^\ddagger$          & \multicolumn{1}{c|}{0.3142$^\ddagger$}          & 0.5628$^\ddagger$          & 0.3564$^\ddagger$          & \multicolumn{1}{c|}{0.2928$^\ddagger$}          & 0.5789$^\ddagger$          & 0.3887$^\ddagger$          & 0.3301$^\ddagger$          \\
\multicolumn{1}{c|}{CoNet}     & 0.3729$^\ddagger$          & 0.2556$^\ddagger$          & \multicolumn{1}{c|}{0.2176$^\ddagger$}    & 0.3887$^\ddagger$          & 0.2658$^\ddagger$          & \multicolumn{1}{c|}{ 0.2183$^\ddagger$}    & 0.5146$^\ddagger$          & 0.3013$^\ddagger$          & \multicolumn{1}{c|}{0.2369$^\ddagger$}          & 0.5380$^\ddagger$           & 0.3140$^\ddagger$          & 0.2229$^\ddagger$          \\
\multicolumn{1}{c|}{sCoNet}     & 0.3773$^\ddagger$          & 0.2594$^\ddagger$          & \multicolumn{1}{c|}{0.2197$^\ddagger$}    & 0.3953$^\ddagger$          & 0.2670$^\ddagger$          & \multicolumn{1}{c|}{0.2080$^\ddagger$}    & 0.5209$^\ddagger$          & 0.3060$^\ddagger$          & \multicolumn{1}{c|}{0.2390$^\ddagger$}          & 0.5411$^\ddagger$           & 0.3148$^\ddagger$          & 0.2283$^\ddagger$          \\
\multicolumn{1}{c|}{ATLRec}     & 0.3771$^\ddagger$          & 0.2758$^\ddagger$          & \multicolumn{1}{c|}{0.2440$^\ddagger$}    & 0.3820$^\ddagger$          & 0.2782$^\ddagger$          & \multicolumn{1}{c|}{0.2435$^\ddagger$}          & 0.5189$^\ddagger$          & 0.3215$^\ddagger$          & \multicolumn{1}{c|}{0.2627$^\ddagger$}          & 0.5198$^\ddagger$          & 0.3225$^\ddagger$          & 0.2514$^\ddagger$          \\
\multicolumn{1}{c|}{DDTCDR}     & 0.3880$^\ddagger$          & 0.2748$^\ddagger$          & \multicolumn{1}{c|}{ 0.2366$^\ddagger$}    & 0.4204$^\ddagger$          & 0.3169$^\ddagger$          & \multicolumn{1}{c|}{0.2804$^\ddagger$}    & 0.5220$^\ddagger$          & 0.3177$^\ddagger$          & \multicolumn{1}{c|}{0.2542$^\ddagger$}          & 0.5421$^\ddagger$           & 0.3563$^\ddagger$          & 0.2962$^\ddagger$          \\
\multicolumn{1}{c|}{DARec}     & \underline{ 0.4589$^\ddagger$}    & \underline{ 0.3349$^\ddagger$}    & \multicolumn{1}{c|}{\underline{0.2950$^\ddagger$}}           & \underline{ 0.4822$^\ddagger$}    & \underline{ 0.3636$^\ddagger$}    & \multicolumn{1}{c|}{\underline{0.3241$^\ddagger$}}          & \underline{ 0.5973$^\ddagger$}    & \underline{ 0.3790$^\ddagger$}    & \multicolumn{1}{c|}{\underline{ 0.3134$^\ddagger$}}    & \underline{ 0.6125$^\ddagger$}    & \underline{ 0.4051$^\ddagger$}    & \underline{ 0.3413$^\ddagger$}    \\ \hline
\multicolumn{1}{c|}{ETL-JRL}  & 0.4869$^\ddagger$          & 0.3629$^\dagger$          & \multicolumn{1}{c|}{0.3210$^\ddagger$}           & 0.5260$^\ddagger$           & 0.4027$^-$          & \multicolumn{1}{c|}{0.3631$^\ddagger$}          & 0.6222$^\ddagger$          & 0.4057$^\ddagger$          & \multicolumn{1}{c|}{0.3387$^\ddagger$}          & 0.6548$^-$          & 0.4422$^\dagger$          & 0.3766$^\ddagger$          \\
\multicolumn{1}{c|}{ETL}      & \textbf{0.4891} & \textbf{0.3632} & \multicolumn{1}{c|}{\textbf{0.3224}} & \textbf{0.5314} & \textbf{0.4037} & \multicolumn{1}{c|}{\textbf{0.3653}} & \textbf{0.6241} & \textbf{0.4076} & \multicolumn{1}{c|}{\textbf{0.3404}} & \textbf{0.6550} & \textbf{0.4442} & \textbf{0.3819} \\ \hline
\multicolumn{1}{c|}{\%Improv.} & 6.58\%          & 8.45\%          & \multicolumn{1}{c|}{9.28\%}          & 10.20\%         & 11.02\%         & \multicolumn{1}{c|}{12.71\%}         & 4.48\%          & 7.54\%          & \multicolumn{1}{c|}{8.61\%}          & 6.93\%          & 9.65\%          & 11.89\%         \\ \hline
\end{tabular}
}}
\vspace{-5pt}
\end{table*}

\begin{table*}[]
\centering
\caption{The overall performance on Music \& Book. The underlined results are the best performance of baselines. Compared to ETL, the t-test results of other baselines are shown in this table. $\ddagger$ means p-value$<$0.01, $\dagger$ indicates p-value$<$0.05 and $-$ means p-value$>$0.05.}
\label{table:overall_amazon3}
\renewcommand{\arraystretch}{1.2}
 \setlength{\tabcolsep}{0.9mm}{ 
  \scalebox{0.85}{
\begin{tabular}{ccccccccccccc}
\hline
\multicolumn{13}{c}{Music \& Book}                                                                                                                                                                                                                                                                           \\ \hline
\multicolumn{1}{c|}{topK}      & \multicolumn{6}{c|}{topK=5}                                                                                                                         & \multicolumn{6}{c}{topK=10}                                                                                                    \\ \hline
\multicolumn{1}{c|}{Domain}    & \multicolumn{3}{c|}{Music}                                               & \multicolumn{3}{c|}{Book}                                                & \multicolumn{3}{c|}{Music}                                               & \multicolumn{3}{c}{Book}                            \\ \hline
\multicolumn{1}{c|}{Metrics}   & HR              & NDCG            & \multicolumn{1}{c|}{MRR}             & HR              & NDCG            & \multicolumn{1}{c|}{MRR}             & HR              & NDCG            & \multicolumn{1}{c|}{MRR}             & HR              & NDCG            & MRR             \\ \hline
\multicolumn{1}{c|}{PMF}       & 0.4213$^\ddagger$          & 0.3138$^\ddagger$          & \multicolumn{1}{c|}{0.2783$^\ddagger$}          & 0.4015$^\ddagger$          & 0.3182$^\ddagger$          & \multicolumn{1}{c|}{0.2889$^\ddagger$}          & 0.5360$^\ddagger$          & 0.3508$^\ddagger$          & \multicolumn{1}{c|}{0.2936$^\ddagger$}          & 0.4992$^\ddagger$          & 0.3480$^\ddagger$          & 0.3009$^\ddagger$          \\
\multicolumn{1}{c|}{CDAE}      & 0.4266$^\ddagger$          & 0.3259$^\ddagger$          & \multicolumn{1}{c|}{0.2839$^\ddagger$}          & 0.4046$^\ddagger$          & 0.3129$^\ddagger$          & \multicolumn{1}{c|}{0.2868$^\ddagger$}          & 0.5471$^\ddagger$          & 0.3615$^\ddagger$          & \multicolumn{1}{c|}{0.3031$^\ddagger$}          & 0.5139$^\ddagger$          & 0.3478$^\ddagger$          & 0.2985$^\ddagger$          \\
\multicolumn{1}{c|}{CFVAE}      & 0.4101$^\ddagger$          & 0.3104$^\ddagger$          & \multicolumn{1}{c|}{0.2718$^\ddagger$}          & 0.3763$^\ddagger$          & 0.2891$^\ddagger$          & \multicolumn{1}{c|}{0.2573$^\ddagger$}          & 0.5342$^\ddagger$          & 0.3488$^\ddagger$          & \multicolumn{1}{c|}{0.2860$^\ddagger$}          & 0.5077$^\ddagger$          & 0.3275$^\ddagger$          & 0.2747$^\ddagger$          \\
\multicolumn{1}{c|}{AAE}       & 0.4302$^\ddagger$          & 0.3326$^\ddagger$          & \multicolumn{1}{c|}{0.3007$^\ddagger$}          & 0.3983$^\ddagger$          & 0.3159$^\ddagger$          & \multicolumn{1}{c|}{0.2852$^\ddagger$}          & 0.5498$^\ddagger$          & 0.3712$^\ddagger$          & \multicolumn{1}{c|}{0.3152$^\ddagger$}          & 0.5121$^\ddagger$          & 0.3491$^\ddagger$          & 0.2992$^\ddagger$          \\ \hline
\multicolumn{1}{c|}{CMF}       & 0.4113$^\ddagger$          & 0.3084$^\ddagger$          & \multicolumn{1}{c|}{0.2748$^\ddagger$}          & 0.4017$^\ddagger$          & 0.3126$^\ddagger$          & \multicolumn{1}{c|}{0.2920$^\ddagger$}          & 0.5280$^\ddagger$          & 0.3468$^\ddagger$          & \multicolumn{1}{c|}{0.2906$^\ddagger$}          & 0.5132$^\ddagger$          & 0.3468$^\ddagger$          & 0.3055$^\ddagger$          \\
\multicolumn{1}{c|}{AAE++}     & 0.4270$^\ddagger$          & 0.3287$^\ddagger$          & \multicolumn{1}{c|}{0.2956$^\ddagger$}          & 0.3996$^\ddagger$          & 0.3200$^\ddagger$          & \multicolumn{1}{c|}{0.2917$^\ddagger$}          & 0.5450$^\ddagger$          & 0.3661$^\ddagger$          & \multicolumn{1}{c|}{0.3110$^\ddagger$}          & 0.5084$^\ddagger$          & 0.3535$^\ddagger$          & 0.3055$^\ddagger$          \\
\multicolumn{1}{c|}{CoNet}     & 0.3380$^\ddagger$          & 0.2235$^\ddagger$          & \multicolumn{1}{c|}{0.2186$^\ddagger$}          & 0.3265$^\ddagger$          & 0.2032$^\ddagger$          & \multicolumn{1}{c|}{0.2061$^\ddagger$}          & 0.4699$^\ddagger$          & 0.2663$^\ddagger$          & \multicolumn{1}{c|}{0.2365$^\ddagger$}          & 0.4505$^\ddagger$          & 0.2452$^\ddagger$          & 0.2419$^\ddagger$          \\
\multicolumn{1}{c|}{sCoNet}     & 0.3508$^\ddagger$          & 0.2370$^\ddagger$          & \multicolumn{1}{c|}{0.2261$^\ddagger$}          & 0.3263$^\ddagger$          & 0.2185$^\ddagger$          & \multicolumn{1}{c|}{0.2297$^\ddagger$}          & 0.4846$^\ddagger$          & 0.2780$^\ddagger$          & \multicolumn{1}{c|}{0.2440$^\ddagger$}          & 0.4490$^\ddagger$          & 0.2590$^\ddagger$          & 0.2460$^\ddagger$          \\
\multicolumn{1}{c|}{ATLRec}     & 0.3540$^\ddagger$          & 0.2512$^\ddagger$          & \multicolumn{1}{c|}{0.2372$^\ddagger$}    & 0.3292$^\ddagger$          & 0.2576$^\ddagger$          & \multicolumn{1}{c|}{0.2376$^\ddagger$}          & 0.4935$^\ddagger$          & 0.2942$^\ddagger$          & \multicolumn{1}{c|}{0.2549$^\ddagger$}          & 0.4522$^\ddagger$          & 0.2773$^\ddagger$          & 0.2529$^\ddagger$          \\
\multicolumn{1}{c|}{DDTCDR}     & 0.3965$^\ddagger$          & 0.3061$^\ddagger$          & \multicolumn{1}{c|}{0.2749$^\ddagger$}          & 0.3689$^\ddagger$          & 0.2992$^\ddagger$          & \multicolumn{1}{c|}{0.2734$^\ddagger$}          & 0.5110$^\ddagger$          & 0.3412$^\ddagger$          & \multicolumn{1}{c|}{0.2879$^\ddagger$}          & 0.4700$^\ddagger$          & 0.3300$^\ddagger$          & 0.2872$^\ddagger$          \\
\multicolumn{1}{c|}{DARec}     & \underline{ 0.4535$^\ddagger$}    & \underline{0.3422$^\ddagger$}    & \multicolumn{1}{c|}{\underline{0.3060$^\ddagger$}}    & \underline{0.4368$^\ddagger$}    & \underline{0.3350$^\ddagger$}    & \multicolumn{1}{c|}{\underline{0.3013$^\ddagger$}}    & \underline{0.5796$^\ddagger$}    & \underline{0.3832$^\ddagger$}    & \multicolumn{1}{c|}{\underline{0.3229$^\ddagger$}}    & \underline{0.5494$^\ddagger$}    & \underline{0.3710$^\ddagger$}    & \underline{0.3161$^\ddagger$}    \\ \hline
\multicolumn{1}{c|}{ETL-JRL}  & 0.4646$^\ddagger$          & 0.3586$^\ddagger$          & \multicolumn{1}{c|}{0.3228$^\ddagger$}          & 0.4458$^\ddagger$          & 0.3389$^\ddagger$          & \multicolumn{1}{c|}{0.3139$^\ddagger$}          & 0.5855$^\ddagger$          & 0.3968$^\ddagger$          & \multicolumn{1}{c|}{0.3385$^\ddagger$}          & 0.5650$^\ddagger$           & 0.3828$^\ddagger$          & 0.3288$^\ddagger$          \\
\multicolumn{1}{c|}{ETL}      & \textbf{0.4686} & \textbf{0.3683} & \multicolumn{1}{c|}{\textbf{0.3282}} & \textbf{0.4496} & \textbf{0.3493} & \multicolumn{1}{c|}{\textbf{0.3155}} & \textbf{0.5942} & \textbf{0.4034} & \multicolumn{1}{c|}{\textbf{0.3444}} & \textbf{0.5669} & \textbf{0.3865} & \textbf{0.3369} \\ \hline
\multicolumn{1}{c|}{\%Improv.} & 3.32\%          & 7.62\%          & \multicolumn{1}{c|}{7.25\%}          & 2.93\%          & 4.26\%          & \multicolumn{1}{c|}{4.71\%}          & 2.51\%          & 5.27\%          & \multicolumn{1}{c|}{6.65\%}          & 3.18\%          & 4.17\%          & 6.58\%          \\ \hline
\end{tabular}
}}
\vspace{-5pt}
\end{table*}
\subsubsection{\textbf{Parameter Settings}}
We implement our ETL with Pytorch on a machine with one Ti-1080 GPU. The embedding size is fixed to 200 for all methods. We optimize all models with Adam optimizer~\cite{kingma2014adam} and the batch size is set as 256. The default Xavier initializer~\cite{glorot2010understanding} is used to initialize all model parameters. For all methods, the dropout ratio is set as 0.5 and the learning rate is 0.001. The number of training epochs is set to 300 which could ensure the convergence for all models.
In ETL, we do not tune hyper-parameter $\eta$ and fix it as 1.0 on the three benchmarks for simplicity. For hyper-parameter $\lambda$, we tune it among [0.1,0.5,1.0,2.0,5.0,10.0] according to the performance on the validation set. Then we obtain $\lambda=5.0$ for Movie \& Book, $\lambda=0.5$ for Movie \& Music and $\lambda=1.0$ for Music \& Book. The codes of PMF, CDAE, CFVAE, AAE and CMF are easily obtained online.
For CoNet, sCoNet, ATLRec and DDTCDR, we directly use the implementation provided by the authors from emails and keep the default settings. Since we do not acquire the codes of DARec from the authors, we implemented them with Pytorch according to details in~\cite{yuan2019darec}.
\begin{figure*}[t]
\centering
\begin{minipage}[t]{0.45\textwidth}
\centering
\includegraphics[width=\textwidth]{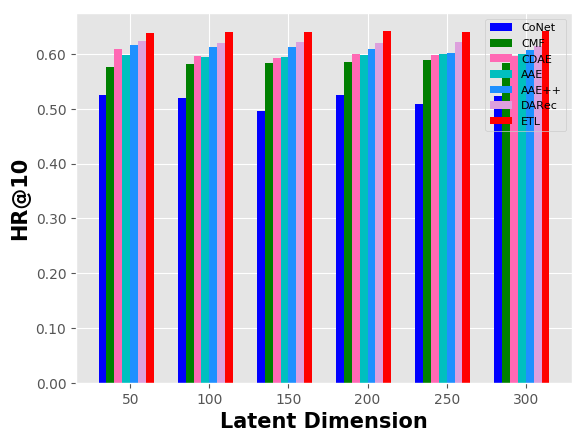}
\vspace{-20pt}
\caption*{(a) \footnotesize{Movie \& Book-Movie}}
\end{minipage}
\begin{minipage}[t]{0.45\textwidth}
\centering
\includegraphics[width=\textwidth]{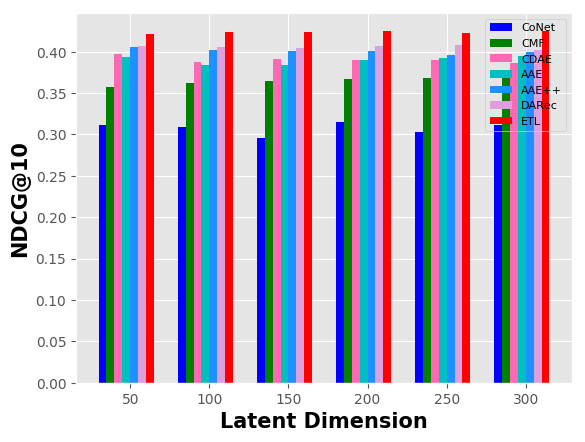}
\vspace{-20pt}
\caption*{(b) \footnotesize{Movie \& Book-Movie}}
\end{minipage} \\
\begin{minipage}[t]{0.45\textwidth}
\centering
\includegraphics[width=\textwidth]{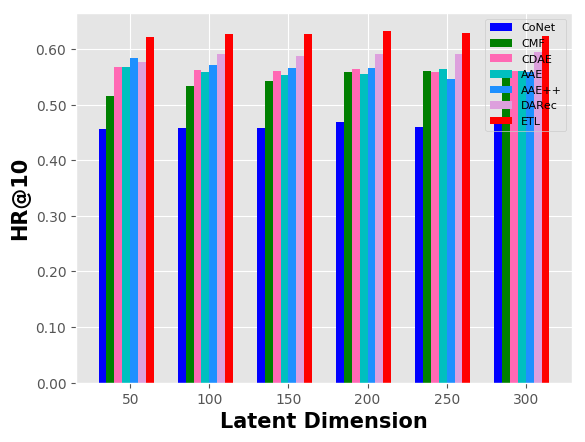}
\vspace{-20pt}
\caption*{(c) \footnotesize{Movie \& Book-Book}}
\end{minipage} 
\begin{minipage}[t]{0.45\textwidth}
\centering
\includegraphics[width=\textwidth]{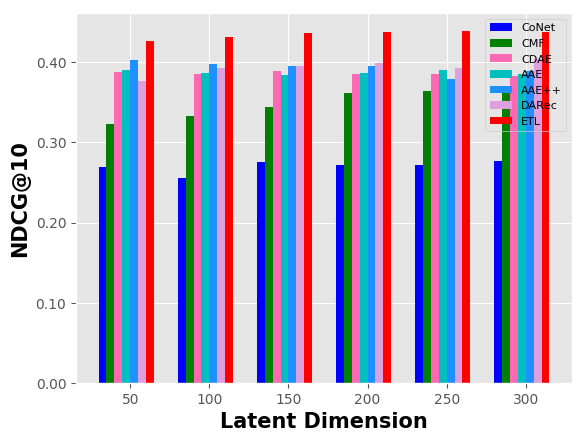}
\vspace{-20pt}
\caption*{(d) \footnotesize{Movie \& Book-Book}}
\end{minipage} \\
\caption{The effects of different latent dimensions on Movie \& Book.}
\label{figure:latent_dim_data1}
\end{figure*}

\begin{figure*}[t]
\centering
\begin{minipage}[t]{0.45\textwidth}
\centering
\includegraphics[width=\textwidth]{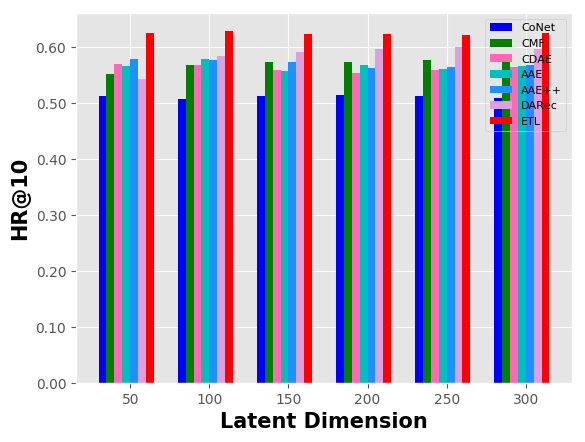}
\vspace{-20pt}
\caption*{(a) \footnotesize{Movie \& Music-Movie}}
\end{minipage}
\begin{minipage}[t]{0.45\textwidth}
\centering
\includegraphics[width=\textwidth]{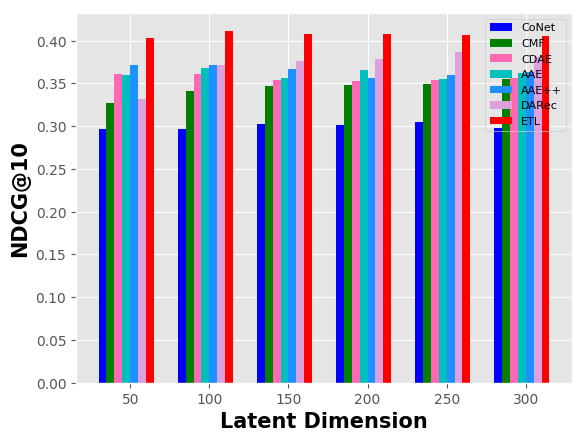}
\vspace{-20pt}
\caption*{(b) \footnotesize{Movie \& Music-Movie}}
\end{minipage} \\
\begin{minipage}[t]{0.45\textwidth}
\centering
\includegraphics[width=\textwidth]{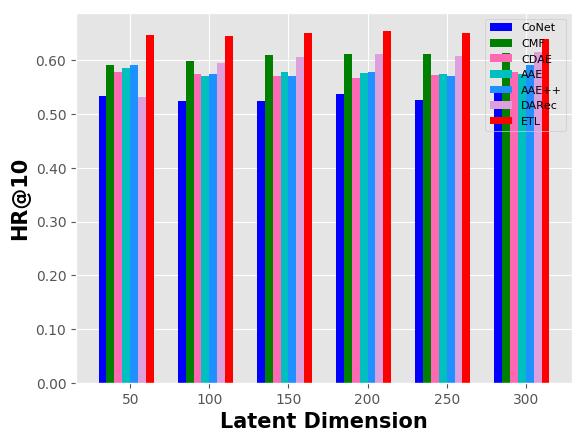}
\vspace{-20pt}
\caption*{(c) \footnotesize{Movie \& Music-Music}}
\end{minipage}
\begin{minipage}[t]{0.45\textwidth}
\centering
\includegraphics[width=\textwidth]{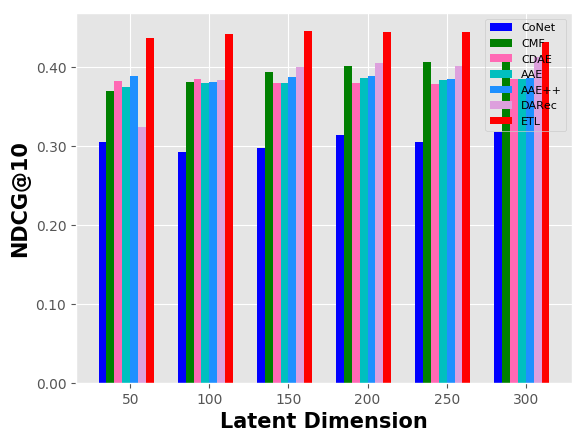}
\vspace{-20pt}
\caption*{(d) \footnotesize{Movie \& Music-Music}}
\end{minipage} \\
\caption{The effects of different latent dimensions on Movie \& Music.}
\label{figure:latent_dim_data2}
\end{figure*}

\begin{figure*}[t]
\centering
\begin{minipage}[t]{0.45\textwidth}
\centering
\includegraphics[width=\textwidth]{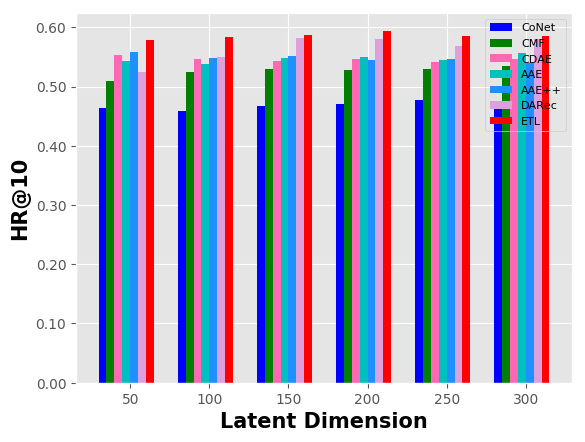}
\vspace{-20pt}
\caption*{(a) \footnotesize{Music \& Book-Music}}
\end{minipage}
\begin{minipage}[t]{0.45\textwidth}
\centering
\includegraphics[width=\textwidth]{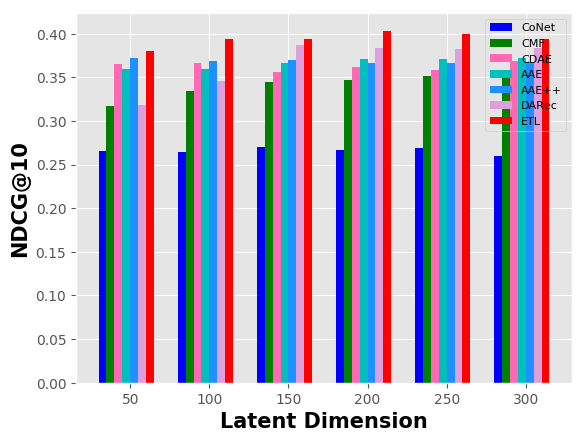}
\vspace{-20pt}
\caption*{(b) \footnotesize{Music \& Book-Music}}
\end{minipage} \\
\begin{minipage}[t]{0.45\textwidth}
\centering
\includegraphics[width=\textwidth]{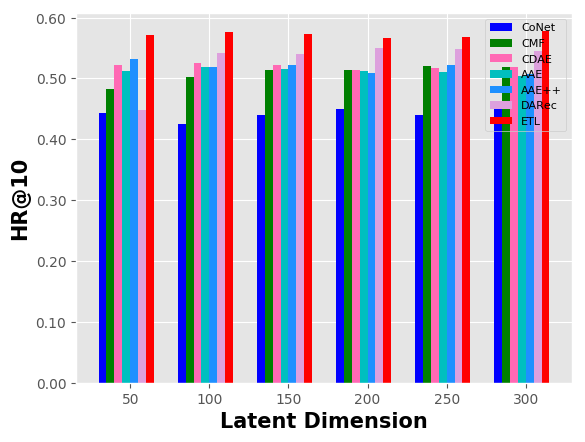}
\vspace{-20pt}
\caption*{(c) \footnotesize{Music \& Book-Book}}
\end{minipage}
\begin{minipage}[t]{0.45\textwidth}
\centering
\includegraphics[width=\textwidth]{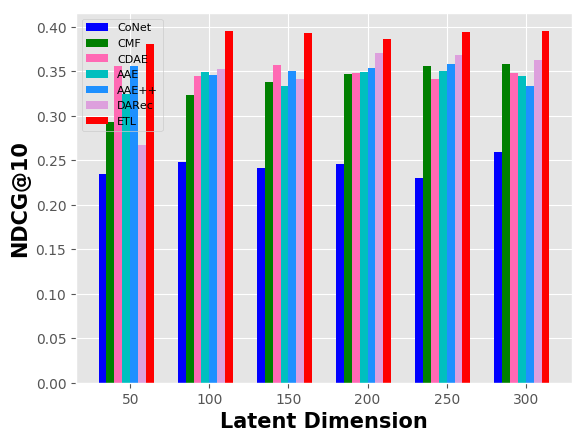}
\vspace{-20pt}
\caption*{(d) \footnotesize{Music \& Book-Book}}
\end{minipage}
\caption{The effects of different latent dimensions on Music \& Book.}
\label{figure:latent_dim_data3}
\end{figure*}
\subsection{Performance Comparison}
\subsubsection{\textbf{Overall Comparison}}\label{sec:overall_performance}
In this evaluation, after we learned the user representations $z_x$ and $z_y$ from the user behaviors in each domain, we use $z_{x}$ and $z_{y}$ to predict the user behaviors by the decoder $D_{x}$ and $D_{y}$, and thus make the recommendation. The performance comparison results are reported in Table~\ref{table:overall_amazon1},\ref{table:overall_amazon2},\ref{table:overall_amazon3}, together with the t-test results. Compared to the most competitive baseline, the percentage of relative improvement (\%Improv.) of ETL is calculated through $100*(v_{ETL}-v_{base})$/$v_{base}$. From these tables, we have the following observations:
\begin{itemize}
 \item ETL consistently yields the best performance on the three datasets. In particular, ETL improves over the most competitive baseline with a 7.54\%, 9.65\% relative gain of NDCG@10 on movie and music domain of Movie \& Music. 
 Compared to DDTCDR, ETL explicitly models the joint distribution of user behaviors across domains by the equivalent transformation assumption and achieves better performance. 
 The superior performance of ETL over AAE++ indicates the importance of cross domain generation in ETL.
 Compared to ATLRec, the proposed ETL contains the cross-generation stream that helps to better learn the user representations in each domain, and shows better performance.
 \item Cross domain based methods generally outperform the single domain based methods, indicating the importance of transferring knowledge across domains in recommendation~\cite{singh2008relational,li2019DDTCDR,yuan2019darec}. In particular, in order to transfer knowledge from other domains, CMF utilizes the linear collective matrix factorization technique while AAE++ and DARec employ various deep learning techniques. Compare to CMF, the proposed ETL consistently shows better performance on different datasets in different sparsity levels. For CoNet, it presents an unsatisfactory performance, because the learning mechanism in CoNet breaks the joint behavior pattern in CDR. 
 \item ETL achieves better performance than ETL-JRL which only has the joint reconstruction loss. The reason for this is that the prior regularization in \textit{ELBO} of Eq.~\ref{eq:ELBO} encourages the preferences to be learned in a specific space with prior knowledge, which benefits the learning process. Moreover, although the original intention of the prior regularization is the joint prior $p(z_{x},z_{y})$, the results show the standard Gaussian distribution in Section~\ref{sec:PRL} also works. This verifies the effectiveness of ETL even with a simple prior.
 \item The improvements of ETL are higher when using Movie as one of the domains. A possible explanation is that the movie domain tends to have more information (\emph{e.g.} background music, prototype book, actor, director, etc.) than other two domains. Compared to previous works that focus on modeling the overlapped features~\cite{singh2008relational,hu2018conet,yuan2019darec,li2019DDTCDR}, ETL has an advantage of capturing more domain-specific information in the movie domain and thus helps the user behavior prediction. 
 \item Considering the t-test results, we can see that ETL presents statistically significant performance compared to the baseline models. It is also worthwhile to point out that in some cases ETL does not have statistically significant performance over ETL-JRL. This is mainly because the prior distribution may be data-dependent and the standard Gaussian distribution in ETL is not the best choice. Nonetheless, it does not serve as a conflict to our main idea. We also discuss the effects of different prior distributions with experiments in Section~\ref{sec:diff_priors}. 
\end{itemize}

\begin{figure*}[t]
\centering
\begin{minipage}[t]{0.45\textwidth}
\centering
\includegraphics[width=\textwidth]{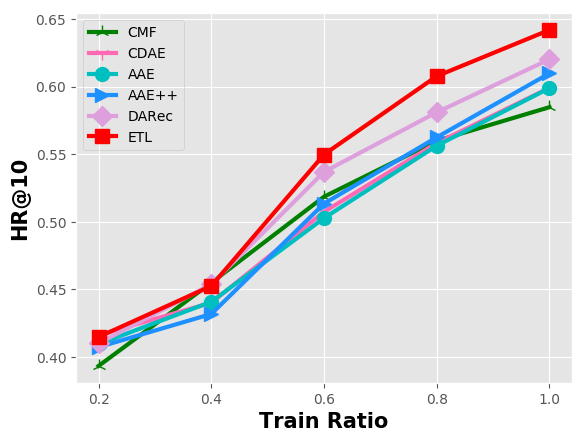}
\vspace{-20pt}
\caption*{(a) \footnotesize{Movie \& Book-Movie}}
\end{minipage}
\begin{minipage}[t]{0.45\textwidth}
\centering
\includegraphics[width=\textwidth]{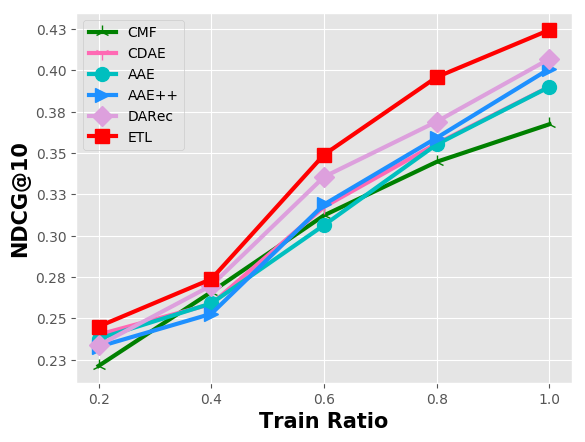}
\vspace{-20pt}
\caption*{(b) \footnotesize{Movie \& Book-Movie}}
\end{minipage} \\
\begin{minipage}[t]{0.45\textwidth}
\centering
\includegraphics[width=\textwidth]{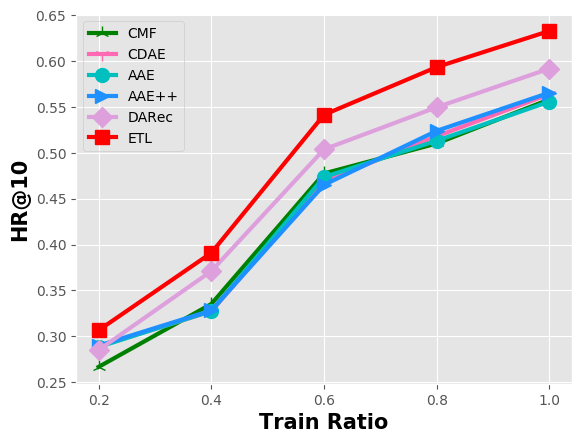}
\vspace{-20pt}
\caption*{(c) \footnotesize{Movie \& Book-Book}}
\end{minipage} 
\begin{minipage}[t]{0.45\textwidth}
\centering
\includegraphics[width=\textwidth]{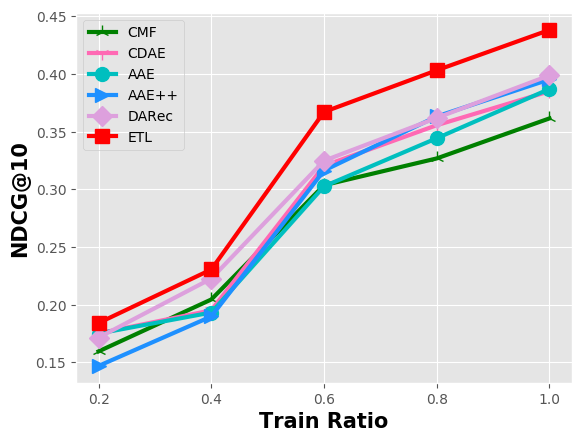}
\vspace{-20pt}
\caption*{(d) \footnotesize{Movie \& Book-Book}}
\end{minipage} \\
\caption{The effects of different sparsity levels on Movie \& Book. Train ratio means the ratio of the original train data.}
\label{figure:sparsity_level_data1}
\end{figure*}

\begin{figure*}[t]
\centering
\begin{minipage}[t]{0.45\textwidth}
\centering
\includegraphics[width=\textwidth]{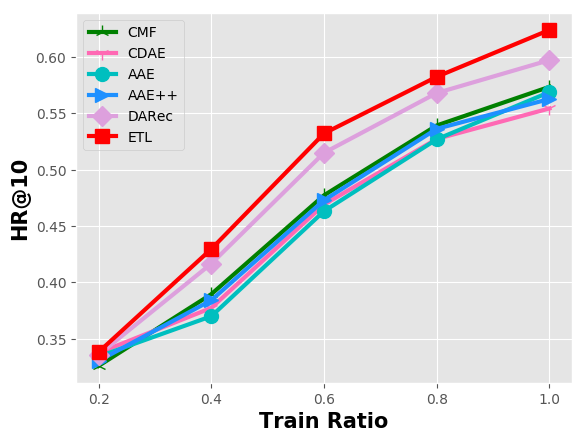}
\vspace{-20pt}
\caption*{(a) \footnotesize{Movie \& Music-Movie}}
\end{minipage}
\begin{minipage}[t]{0.45\textwidth}
\centering
\includegraphics[width=\textwidth]{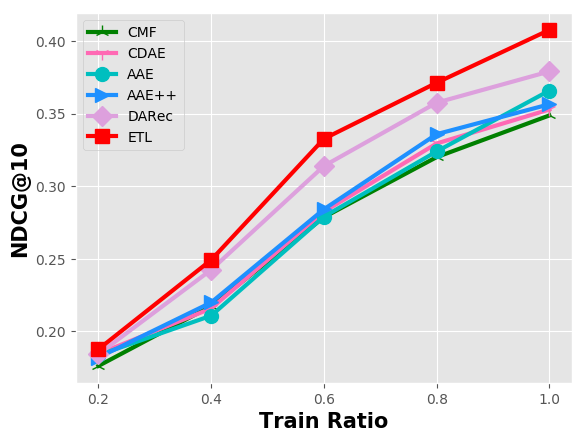}
\vspace{-20pt}
\caption*{(b) \footnotesize{Movie \& Music-Movie}}
\end{minipage} \\
\begin{minipage}[t]{0.45\textwidth}
\centering
\includegraphics[width=\textwidth]{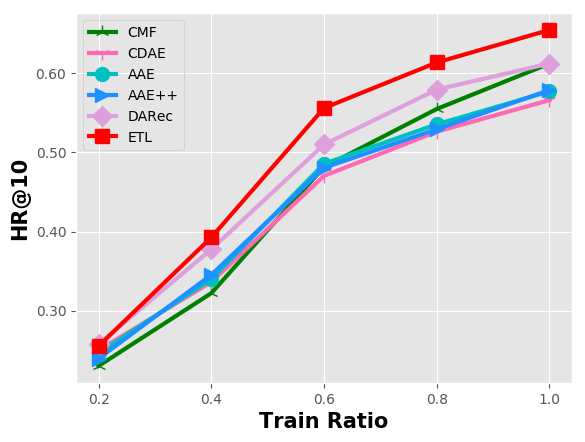}
\vspace{-20pt}
\caption*{(c) \footnotesize{Movie \& Music-Music}}
\end{minipage}
\begin{minipage}[t]{0.45\textwidth}
\centering
\includegraphics[width=\textwidth]{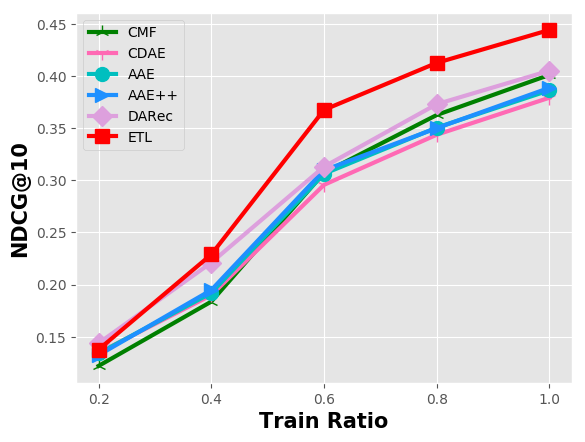}
\vspace{-20pt}
\caption*{(d) \footnotesize{Movie \& Music-Music}}
\end{minipage} \\
\caption{The effects of different sparsity levels on Movie \& Music. Train ratio means the ratio of the original train data.}
\label{figure:sparsity_level_data2}
\end{figure*}

\begin{figure*}[t]
\centering
\begin{minipage}[t]{0.45\textwidth}
\centering
\includegraphics[width=\textwidth]{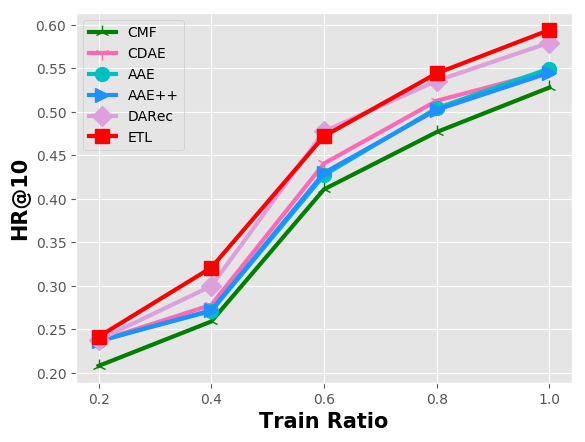}
\vspace{-20pt}
\caption*{(a) \footnotesize{Music \& Book-Music}}
\end{minipage}
\begin{minipage}[t]{0.45\textwidth}
\centering
\includegraphics[width=\textwidth]{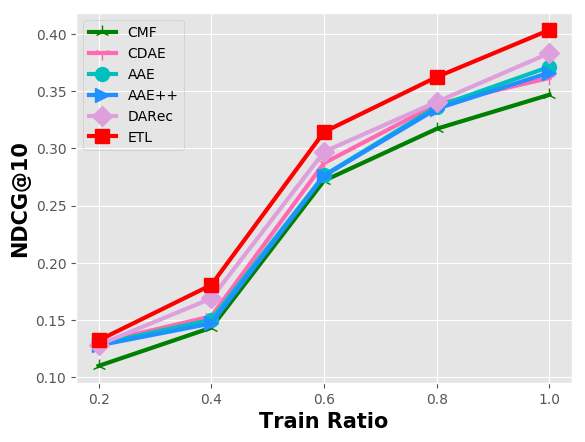}
\vspace{-20pt}
\caption*{(b) \footnotesize{Music \& Book-Music}}
\end{minipage} \\
\begin{minipage}[t]{0.45\textwidth}
\centering
\includegraphics[width=\textwidth]{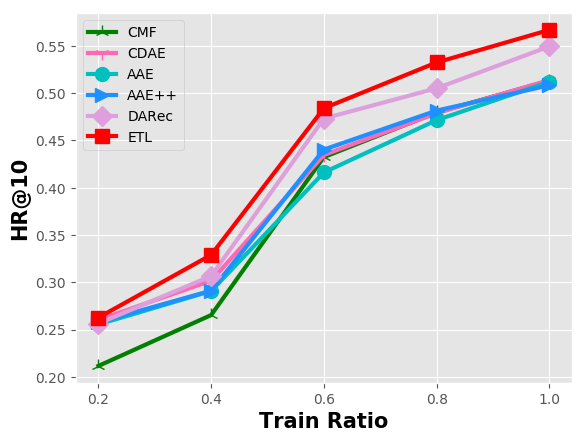}
\vspace{-20pt}
\caption*{(c) \footnotesize{Music \& Book-Book}}
\end{minipage}
\begin{minipage}[t]{0.45\textwidth}
\centering
\includegraphics[width=\textwidth]{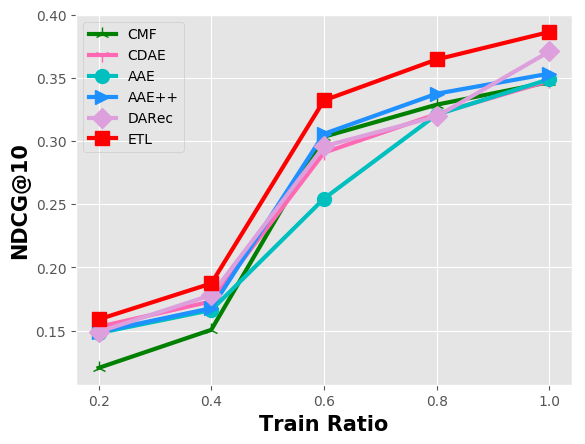}
\vspace{-20pt}
\caption*{(d) \footnotesize{Music \& Book-Book}}
\end{minipage}
\caption{The effects of different sparsity levels on Music \& Book. Train ratio means the ratio of the original train data.}
\label{figure:sparsity_level_data3}
\end{figure*}

\subsubsection{\textbf{Different Latent Dimensions}}
The latent dimension is an important factor that accounts for the recommendation performance of different methods. 
We thus investigate the impact of different latent dimensions for different methods. 
To be specific, we fix the other factors of all methods and allow the latent dimension $d$ to range in [50,100,150,200,250,300]. The results on three datasets are shown in Figure~\ref{figure:latent_dim_data1}, Figure~\ref{figure:latent_dim_data2} and Figure~\ref{figure:latent_dim_data3}.  From these figures, we summarize that:
\begin{itemize}
\item Compared with other methods, ETL consistently achieves the best performance on almost every latent dimension. ETL is robust to the change of latent dimension according to the slight change of performance. This verifies the significance of modeling both features, which helps to robustly transfer knowledge with different latent dimensions.
\item It is worthwhile to point out that DARec does not perform robustly with the change of latent dimensions. The main reason is that DARec involves the pretraining of AutoRec~\cite{sedhain2015autorec} as the base model, which could accumulate noise with a bad latent dimension because of the two-step scheme.
\end{itemize}

\subsubsection{\textbf{Different Sparsity Levels}}
Since sparsity is an important problem in recommendation systems, it is necessary to investigate whether ETL can still perform better than other methods under more sparse conditions.
To this end, we vary sparsity levels of the training data to investigate the method's corresponding performance. 
In particular, fixing the validation and test set, we randomly sample a ratio ranging in [20\%,40\%,60\%,80\%,100\%] of the original train data as the new train data for different sparsity levels. The results are shown in Figure~\ref{figure:sparsity_level_data1}, Figure~\ref{figure:sparsity_level_data2} and Figure~\ref{figure:sparsity_level_data3}. According to these figures, we can see that:
\begin{itemize}
    \item It is obvious that ETL performs better than other methods on almost all sparsity levels. 
    We also observe that ETL has less improvement over other counterparts when the data are extremely sparse. One possible reason is that ETL involves the parameterized equivalent transformation that needs necessary data to exert its performance. Too sparse data may limit the training of the transformation, which may be a limitation of ETL and could be explored in the future.
    \item The extreme sparse cases would cause deterioration to the recommendation performance. As shown in Figure~\ref{figure:sparsity_level_data1}, Figure~\ref{figure:sparsity_level_data2} and Figure~\ref{figure:sparsity_level_data3}, all methods would have a decrease when training data are less. In the extreme sparse case 0.2, the gap among most CDR methods is not obvious, because the observed data are too less to train a reliable model with 0.2 train ratio. For example, with 0.2 train ratio, we only have 0.01\% observed interactions on Movie \& Book-Book.
\end{itemize}
\begin{figure}[t]
\centering
\begin{minipage}[t]{0.32\textwidth}
\centering
\includegraphics[width=\textwidth]{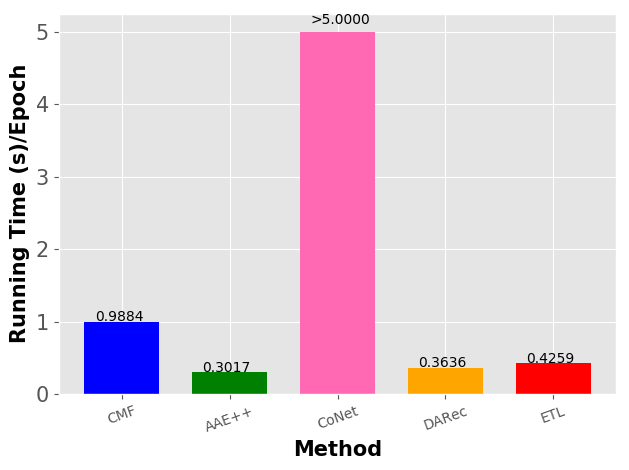}
\vspace{-20pt}
\caption*{(a) \footnotesize{Movie \& Book}}
\end{minipage}
\begin{minipage}[t]{0.32\textwidth}
\centering
\includegraphics[width=\textwidth]{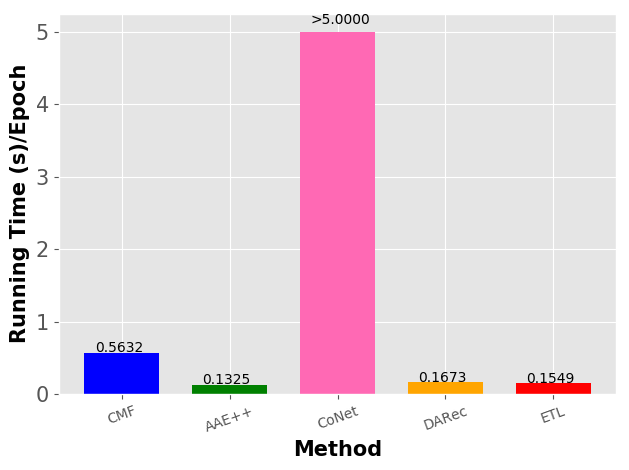}
\vspace{-20pt}
\caption*{(b) \footnotesize{Movie \& Music}}
\end{minipage} 
\begin{minipage}[t]{0.32\textwidth}
\centering
\includegraphics[width=\textwidth]{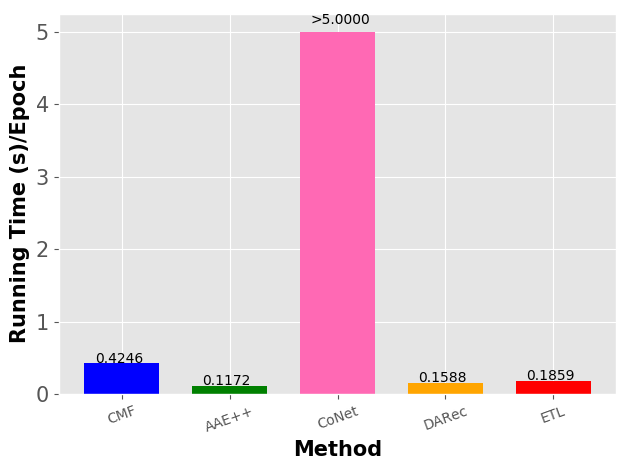}
\vspace{-20pt}
\caption*{(c) \footnotesize{Music \& Book}}
\end{minipage} 
\caption{The empirical running time in each epoch of different methods. }
\label{figure:running_time}
\end{figure}
\subsubsection{\textbf{Empirical Running Time Analysis}}
To investigate the time complexity, we conduct an experiment to compare the empirical running time of each epoch for different models. We conduct the experiments 10 times on the same machine with one Ti-1080 GPU. The mean value of running time per epoch is reported in Figure~\ref{figure:running_time}. 

From Figure~\ref{figure:running_time}, we can see that: (1) CoNet costs the most time because it involves the pretraining of a single domain and stacks multiple cross connection units to model the interactions of different domains. (2) Compared to AAE++ and DARec, ETL takes the running time in the same level and achieves better recommendation performance. This also verifies the time efficiency of the proposed model. (3) The matrix factorization based method CMF costs more time than auto-encoding based methods (\emph{e.g.} AAE++, DARec and ETL). This is because with the same batch size, CMF takes user-item interaction pairs for training, while auto-encoding based methods can take all behaviors of batch users. 

\subsection{Analysis of Learned User Preferences}
As we mentioned before, the equivalent transformation enables ETL to better capture both the domain-specific and overlapped features for CDR. In order to verify this, we design two experiments in this section. 
Note that our model mainly concentrates on the user-item interactions, specific instances about the overlapped and domain-specific features is beyond the interest of this study. 

\subsubsection{\textbf{Overlapped Features of User Preferences}}
In this experiment, we make a hypothesis that the preferences (representations) across domains belonging to the same user usually would have more overlaps than those belonging to different users in CDR.
If the overlaps are well captured, there will be an obvious difference between the embedding pair of the same user and that of different users, which means it can be formulated as a binary classification problem. This intuition is similar to that in some multi-view translation works~\cite{zhu2017toward,isola2017image,guo2019canonical,9229522,shi2020multiview} that use a binary classifier to measure the distribution-level distance between the joint distribution $q(z_{x},z_{y})$ and $q(z_{x},z_{y'})$ where $z_{x},z_{y}$ belong to the same instance and $z_{x},z_{y'}$ belong to different instances, and judge whether the two embeddings have consistent features. Thereby, we employ a binary classifier here to verify whether ETL can better capture the overlapped features of user preferences across domains.

In particular, we denote $Z_{x}\in \mathbb{R}^{N\times d}$ and $Z_{y}\in \mathbb{R}^{N\times d}$ as the user embedding matrix for $\mathcal{X}$ and $\mathcal{Y}$ domain, respectively. In this experiment, we hold a rule where representations (\emph{i.e.} $Z_{xi}$ and $Z_{yi}$) belonging to the same user are the paired sample and representations (\emph{i.e.} $Z_{xi}$ and $Z_{yj}$ with $j\neq i$) belonging to different users are the unpaired sample. We obtain the classification data for this experiment by constructing one paired sample and one random unpaired sample for each user.
Then, we train a two-layer MLP classifier to perform the binary classification task with $(Z_{xi},Z_{yi})$ as label 1 and $(Z_{xi},Z_{yj})$ as label 0. In this experiment, the concatenation operation is employed between $Z_{x}$ and $Z_{y}$ and we follow the common 6(train)-2(validation)-2(test) setting for classification. The experiment is conducted 10 times and we report the mean value as the results\footnote{CMF and CoNet both learn one low-dimensional embedding for each user across domains, which makes them not suitable for this experiment here.}. The results are shown in Table~\ref{table:learned_joint_preference}. There are two key observations:
\vspace{-3pt}
\begin{itemize}
    \item The proposed ETL yields the best classification performance. 
    This indicates that ETL has better ability to discriminate embeddings that belongs to the same user. In other words, ETL can better capture the overlapped features of the same user and distinguish the user from others. It is also worthwhile to mention that the result accuracy is consistently lower for Music \& Book combo since the correlations between Music and Book domain are not as high as other two domains. As we analyzed before, with the idea of shared-user representation, the CDR model may converge to a comprised solution where both the overlapped and domain-specific features are not well-captured. In contrast, the equivalent transformation in ETL allows the flexibility for user representations in each domain, and can  also better learn the overlapped features by the transformed user representations. Learning the overlapped features and domain-specific features is not a conflict for better recommendation performance as long as we find an effective way to do the feature alignment. Through the results, we can see that ETL indeed better captures the correlations across domains.
    \item Two interesting phenomenons are observed in Table~\ref{table:learned_joint_preference}.
    First, the single domain based methods (PMF, CDAE, AAE) have worse AUC compared with the cross domain based methods (AAE++, DARec and ETL). Second, considering the recommendation performance in Table~\ref{table:overall_amazon1},\ref{table:overall_amazon2},\ref{table:overall_amazon3}, it seems that the recommendation accuracy has a positive correlation with the classification AUC, which implies that a model with better classification performance tends to have better recommendation accuracy. These two phenomenons emphasize our motivation of learning $z_{x}$ and $z_{y}$ in CDR with equivalent transformation that helps to learn the overlapped features. It is quite different from the idea of other works that learns those features by fully aligning $z_{x}$ and $z_{y}$, where the alignment can be affected by the user behavior prediction loss in each domain.
\end{itemize}
\begin{table}[]
\centering
\caption{The binary classification results with AUC. This experiment is designed to verify that the proposed ETL can better learn the overlapped features of user preferences in CDR.}
\label{table:learned_joint_preference}
\renewcommand{\arraystretch}{1.2}
 \setlength{\tabcolsep}{1.0mm}{ 
  \scalebox{1.0}{
\begin{tabular}{c|c|c|c} 
\hline
      & \begin{tabular}[c]{@{}c@{}}Movie \& Book\end{tabular} & \begin{tabular}[c]{@{}c@{}} Movie \& Music\end{tabular} & \begin{tabular}[c]{@{}c@{}} Music \& Book\end{tabular} \\ \hline
PMF   & 0.6571                                                           & 0.7322                                                            & 0.5709                                                           \\ \hline
CDAE  & 0.6796                                                           & 0.7267                                                            & 0.6195                                                           \\ \hline
AAE   & 0.7128                                                           & 0.7734                                                            & 0.6516                                                           \\ \hline
AAE++ & 0.7253                                                           & 0.7777                                                            & 0.6565                                                           \\ \hline
DARec & 0.7460                                                            & 0.8203                                                            & 0.6934                                                           \\ \hline
ETL  & \textbf{0.9160}                                                   & \textbf{0.9574}                                                   & \textbf{0.8826}                                                  \\ \hline
\end{tabular}
}}
\end{table}

\begin{figure*}[t]
\centering
\begin{minipage}[t]{0.45\textwidth}
\centering
\includegraphics[width=\textwidth]{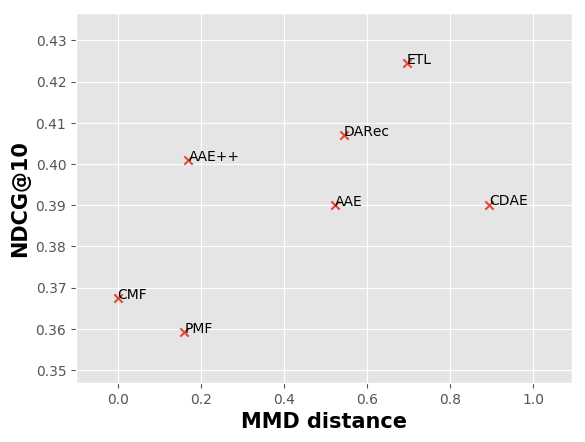}
\vspace{-20pt}
\caption*{(a) \footnotesize{Movie \& Book-Movie}}
\end{minipage}
\begin{minipage}[t]{0.45\textwidth}
\centering
\includegraphics[width=\textwidth]{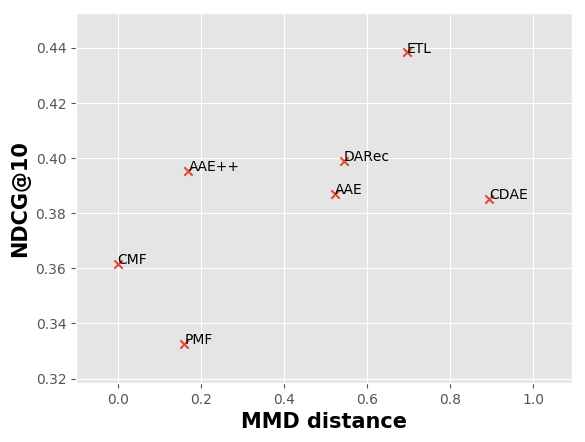}
\vspace{-20pt}
\caption*{(b) \footnotesize{Movie \& Book-Book}}
\end{minipage} \\
\begin{minipage}[t]{0.45\textwidth}
\centering
\includegraphics[width=\textwidth]{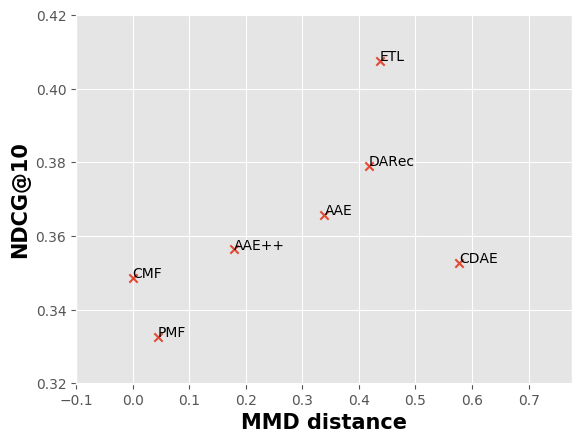}
\vspace{-20pt}
\caption*{(c) \footnotesize{Movie \& Music-Movie}}
\end{minipage}
\begin{minipage}[t]{0.45\textwidth}
\centering
\includegraphics[width=\textwidth]{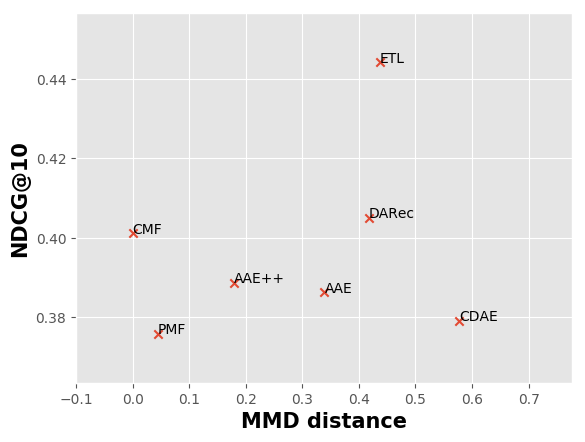}
\vspace{-20pt}
\caption*{(d) \footnotesize{Movie \& Music-Music}}
\end{minipage} \\
\begin{minipage}[t]{0.45\textwidth}
\centering
\includegraphics[width=\textwidth]{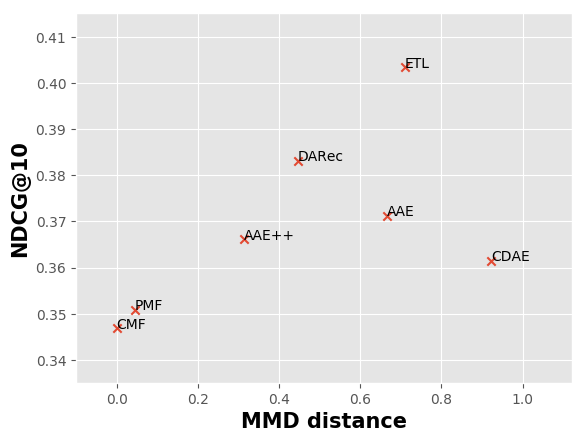}
\vspace{-20pt}
\caption*{(e) \footnotesize{Music \& Book-Music}}
\end{minipage}
\begin{minipage}[t]{0.45\textwidth}
\centering
\includegraphics[width=\textwidth]{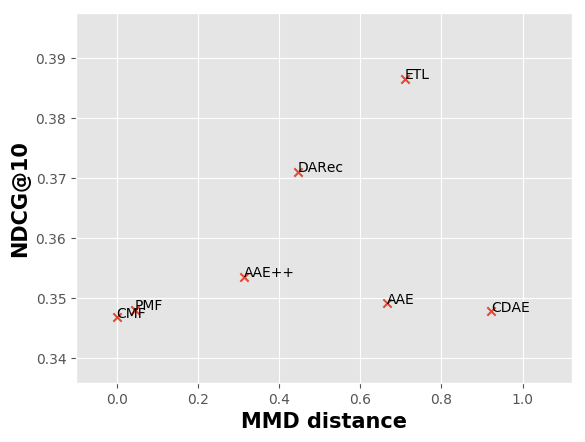}
\vspace{-20pt}
\caption*{(f) \footnotesize{Music \& Book-Book}}
\end{minipage} \\
\caption{The results of model performance and MMD distance on three benchmarks. This experiment is designed to show that ETL has the capacity to learn the domain-specific features of user preferences.}
\label{figure:complete_distribution_level}
\end{figure*}
\subsubsection{\textbf{Domain-Specific Features of User Preferences}}
Assuming the overlapped features are well captured, a CDR model can further boost the recommendation by learning the domain-specific features. 
The idea of shared-user representation expects to align $z_{x}$ and $z_{y}$, and thus leads to low distance between $q(z_{x}|x)$ and $q(z_{y}|y)$ in two domains. In contrast, when the domain-specific features are captured, $q(z_{x}|x)$ and $q(z_{y}|y)$ are expected to have a larger distribution distance.

We thus can verify whether ETL is able to learn the domain-specific features by measuring the distance between $q(z_{x}|x)$ and $q(z_{y}|y)$. 
In particular, for different methods, we use the $z_{x}$ and $z_{y}$ learned by different methods to calculate the 
Maximum Mean Discrepancy (MMD) distance~\cite{gretton2012kernel}. 
Let $\mathscr{F}$ be a class of functions $f:\mathscr{X} \rightarrow \mathbb{R}$,
then the MMD distance is defined as:
\begin{align}
    MMD[\mathscr{F},q(z_{x}|x),q(z_{y}|y)]=\sup_{f\in \mathscr{F}}(\mathbb{E}_{z_{x}}[f(z_{x})]-\mathbb{E}_{z_{y}}[f(z_{y})]) \nonumber
\end{align}
When given $n$ samples from $q(z_{x}|x)$ and $m$ samples from $q(z_{y}|y)$, we can empirically estimate the MMD distance with the following equation\footnote{https://en.wikipedia.org/wiki/Kernel\_embedding\_of\_distributions}:
\begin{align}
\widehat{MMD}=\frac{1}{n^2}\sum_{i=1}^{n}\sum_{j=1}^{n}k(z_{x}^{i},z_{x}^{j})+\frac{1}{m^2}\sum_{i=1}^{m}\sum_{j=1}^{m}k(z_{y}^{i},z_{y}^{j})-\frac{2}{nm}\sum_{i=1}^{n}\sum_{j=1}^{m} k(z_{x}^{i},z_{y}^{j}) \nonumber
\end{align}
where $z_{x}^{i~or~j}$ and $z_{y}^{i~or~j}$ are embedding vectors and $k(\cdot,\cdot)$ denotes the kernel function. In our case, we use the widely used Gaussian radial basis function (RBF) kernel that is defined as $k(x,x')=exp(-\frac{1}{2\sigma^2\|x-x'\|^2})$. Following recent works~\cite{NIPS2017_dfd7468a}, we use multiple kernels of different $\sigma$ and then sum them as the final kernel function\footnote{https://github.com/OctoberChang/MMD-GAN/blob/master/mmd.py}.
The results of MMD distance and recommendation performance on three benchmarks are shown in Figure~\ref{figure:complete_distribution_level}. From this figure, we have the following observations:
\begin{itemize}
    \item Based on the well-captured overlapped features, ETL can further learn the domain-specific features for better recommendation performance and thus has larger MMD distance than other CDR methods\footnote{CMF and CoNet both share exactly the same user representation and have zero MMD distance.}.
    Unlike the idea of shared-user representation that compresses domain-specific features, the idea of equivalent transformation enables ETL to have the capacity of capturing the domain-specific variations as well as the correlations across domains.
    \item It is worthwhile to point out that large distance between $q(z_{x}|x)$ and $q(z_{y}|y)$ does not mean better recommendation performance. The recommendation performance in CDR relies on both the overlapped features and the domain-specific features. In this context, we can understand why some single domain methods (\emph{e.g.} CDAE and AAE) have larger MMD distance but achieve poor recommendation performance.
    \item It is also worthwhile to mention that learning better overlapped features in ETL does not mean the distance between $z_{x}$ and $z_{y}$ should be lower. ETL learns the overlapped features between $z_{x}$ and $z_{y}$ by an equivalent transformation, which means the feature alignment is conducted between $z_{x}$ and $z_{y}W$ (\emph{resp.} $z_{y}$ and $z_{x}W^T$) and thus has no conflict of having large distance between $z_{x}$ and $z_{y}$ when capturing the domain-specific features.
\end{itemize}

\subsection{Ablation Study}
\begin{table}[]
\caption{The definitions of different transformations. Trans1 and Trans2 are both not equivalent transformation. Trans3 is one simple equivalent transformation. Trans4 is our extended non-linear equivalent transformation from Trans3.}
\label{table:trans_styles}
\renewcommand{\arraystretch}{1.0}
 \setlength{\tabcolsep}{1.2mm}{ 
  \scalebox{0.9}{
\begin{tabular}{c|c|c}
\hline
Trans & Formulation & Type                                                                     \\ \hline
Trans1         & \begin{tabular}[c]{@{}c@{}}$z_{y\rightarrow x}=z_{y}W_{x}$,
$z_{x\rightarrow y}=z_{x}W_{y}$
\end{tabular}           & \begin{tabular}[c]{@{}c@{}}\emph{unconstrained}\\ (linear)\end{tabular}      \\ \hline
Trans2         & \begin{tabular}[c]{@{}c@{}}$z_{y\rightarrow x}=\sigma(z_{y}W_{x}^{1})W_{x}^{2}$,
$z_{x\rightarrow y}=\sigma(z_{x}W_{y}^{1})W_{y}^{2}$
\end{tabular}           & \begin{tabular}[c]{@{}c@{}}\emph{unconstrained} \\ (non-linear)\end{tabular} \\ \hline
Trans3         & \begin{tabular}[c]{@{}c@{}}$z_{y\rightarrow x}=z_{y}W_{x}$,
$z_{x\rightarrow y}=z_{x}W_{y}$\\
$s.t.~\min ||z_{x}-z_{x\rightarrow y}W_{x}||_{F}^{1}$\\
$s.t.~\min ||z_{y}-z_{y\rightarrow x}W_{y}||_{F}^{1}$
\end{tabular}           & \begin{tabular}[c]{@{}c@{}}\emph{constrained} \\(linear, equivalent)\end{tabular}         \\ \hline
Trans4         & \begin{tabular}[c]{@{}c@{}}$z_{y\rightarrow x}=\sigma(z_{y}W_{x}^{1})W_{x}^{2}$,
$z_{x\rightarrow y}=\sigma(z_{x}W_{y}^{1})W_{y}^{2}$ \\
$s.t.~\min ||z_{x}-\sigma(z_{x\rightarrow y}W_{x}^{1})W_{x}^{2}||_{F}^{1}$\\
$s.t.~\min ||z_{y}-\sigma(z_{y\rightarrow x}W_{y}^{1})W_{y}^{2}||_{F}^{1}$
\end{tabular}           & \begin{tabular}[c]{@{}c@{}}\emph{constrained} \\ (non-linear, equivalent)\end{tabular}     \\ \hline
\end{tabular}
}}
\end{table}
\begin{figure}[t]
\centering
\begin{minipage}[t]{0.45\textwidth}
\centering
\includegraphics[width=\textwidth]{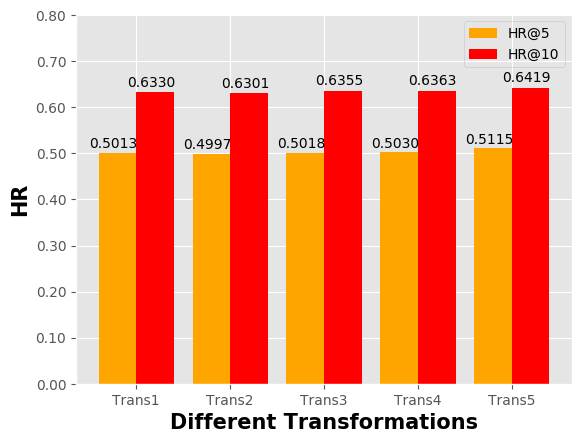}
\vspace{-20pt}
\caption*{(a) \footnotesize{Movie-HR}}
\end{minipage}
\begin{minipage}[t]{0.45\textwidth}
\centering
\includegraphics[width=\textwidth]{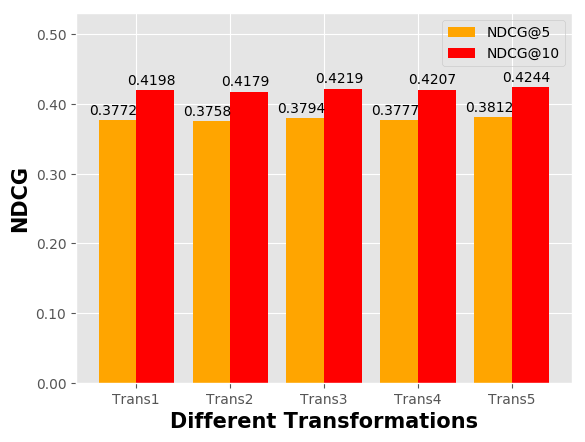}
\vspace{-20pt}
\caption*{(b) \footnotesize{Movie-NDCG}}
\end{minipage} \\
\begin{minipage}[t]{0.45\textwidth}
\centering
\includegraphics[width=\textwidth]{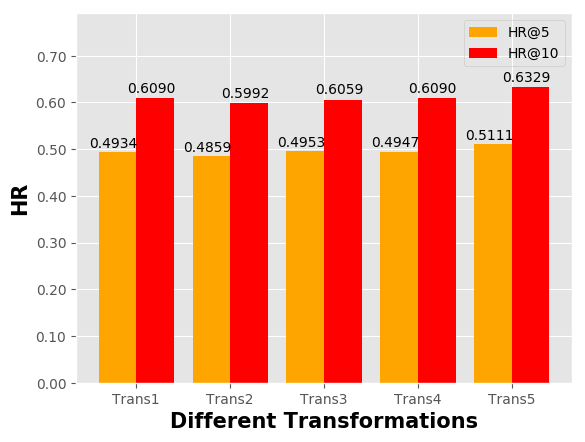}
\vspace{-20pt}
\caption*{(c) \footnotesize{Book-HR}}
\end{minipage} 
\begin{minipage}[t]{0.45\textwidth}
\centering
\includegraphics[width=\textwidth]{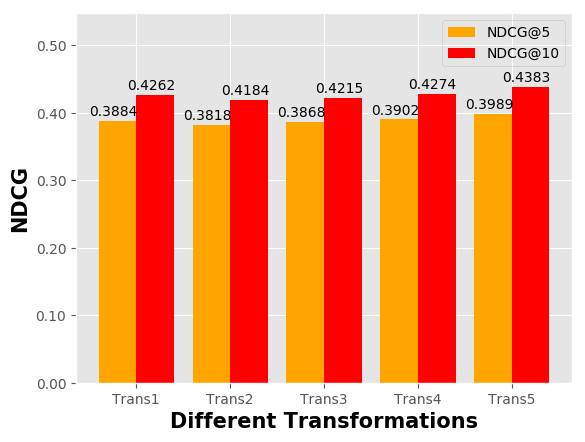}
\vspace{-20pt}
\caption*{(d) \footnotesize{Book-NDCG}}
\end{minipage} \\
\caption{The effect of different transformations on Movie \& Book. Note that we show the results on Movie \& Book here as an example to illustrate our idea. Results on other two benchmarks follow similar pattern. }
\label{figure:diff_trans}
\end{figure}

\subsubsection{\textbf{The Effects of Different Transformations}}
\label{sec:diff_trans}
In case that we adopt the orthogonal transformation for the equivalent transformation, it is curious to explore what are the effects of other transformations. We here conduct an experiment with other 4 different transformations for ETL.

The results on Movie \& Book are shown in Figure~\ref{figure:diff_trans}, where Trans5 is the used orthogonal transformation and Trans1$\sim$4 indicate other 4 different transformations that are defined in Table~\ref{table:trans_styles}.
From Table~\ref{table:trans_styles} and Figure~\ref{figure:diff_trans}, we can see:
\begin{itemize}
    \item The orthogonal transformation (Trans5) outperforms other 4 transformations. This is reasonable since only Trans5 is the equivalent one as well as does not introduce spurious correlations between users after knowledge transfer.
    \item It is worthwhile to notice that the equivalent transformations (Trans3$\sim$5) present better performance compared to the non-equivalent ones (Trans1$\sim$2). This verifies the correctness of our ET assumption that encourages ETL to learn better coverage of user preferences. The \emph{unconstrained} ones deteriorate the recommendation performance since they are too flexible and may easily be over-fitting on the sparse data in recommendation. Moreover, even with other equivalent transformations (Trans3$\sim$4), ETL could still perform better than DARec according to Table~\ref{table:overall_amazon1}$\sim$\ref{table:overall_amazon3}.
\end{itemize}

\begin{table*}[]
\caption{The effects of different priors on three benchmarks. Uniform indicates samples from $U(0,1)$, Laplace indicates samples from $L(0,1)$, Gaussian means samples from $\mathcal{N}(0,1)$ and
MVGaussian means samples from multi-variate Gaussian $MVG=\mathcal{N}(0,1)+\mathcal{N}(3,1)$. For MVGaussian, we simply take the summarization of two independent Gaussian here and the mean value of the second Gaussian is set as 3 in order to form multiple peaks according to the 3-$\sigma$ principle in Gaussian distribution.} 
\label{table:diff_priors}
\renewcommand{\arraystretch}{1.2}
 \setlength{\tabcolsep}{0.5mm}{ 
  \scalebox{0.9}{
\begin{tabular}{ccccccccccccc}
\hline
\multicolumn{13}{c}{Movie and Book}                                                                                                                                                                                                                                                                                                                                                                                                                                                                                                                                                      \\ \hline
\multicolumn{1}{c|}{topK}       & \multicolumn{6}{c|}{topK=5}                                                                                                                                                                                                                                                          & \multicolumn{6}{c}{topK=10}                                                                                                                                                                                                                                     \\ \hline
\multicolumn{1}{c|}{Domain}     & \multicolumn{3}{c|}{Movie}                                                                                                                    & \multicolumn{3}{c|}{Book}                                                                                                            & \multicolumn{3}{c|}{Movie}                                                                                                                    & \multicolumn{3}{c}{Book}                                                                                        \\ \hline
\multicolumn{1}{c|}{Metrics}    & HR                                     & NDCG                                   & \multicolumn{1}{c|}{MRR}                                    & HR                                     & NDCG                          & \multicolumn{1}{c|}{MRR}                                    & HR                                     & NDCG                                   & \multicolumn{1}{c|}{MRR}                                    & HR                                     & NDCG                          & MRR                                    \\ \hline
\multicolumn{1}{c|}{Uniform}    & 0.5030                                 & 0.3761                                 & \multicolumn{1}{c|}{0.3348}                                 & 0.4976                                 & 0.3761                        & \multicolumn{1}{c|}{0.3368}                                 & 0.6341                                 & 0.4185                                 & \multicolumn{1}{c|}{0.3527}                                 & 0.6211                                 & 0.4167                        & 0.3536                                 \\
\multicolumn{1}{c|}{Laplace}    & 0.5095                                 & \textbf{0.3826}                        & \multicolumn{1}{c|}{0.3406}                                 & \textbf{0.5149}                        & \textbf{0.4052}               & \multicolumn{1}{c|}{0.3637}                                 & 0.6418                                 & \textbf{0.4254}                        & \multicolumn{1}{c|}{0.3583}                                 & 0.6320                                 & \textbf{0.4437}               & 0.3792                                 \\
\multicolumn{1}{c|}{MVGaussian} & 0.5044                                 & 0.3770                                 & \multicolumn{1}{c|}{0.3349}                                 & 0.4960                                 & 0.3788                        & \multicolumn{1}{c|}{0.3399}                                 & 0.6398                                 & 0.4208                                 & \multicolumn{1}{c|}{0.3529}                                 & 0.6191                                 & 0.4186                        & 0.3564                                 \\
\multicolumn{1}{c|}{Gaussian}   & {\color[HTML]{333333} \textbf{0.5115}} & {\color[HTML]{333333} 0.3812}          & \multicolumn{1}{c|}{{\color[HTML]{333333} \textbf{0.3431}}} & {\color[HTML]{333333} 0.5111}          & {\color[HTML]{333333} 0.3989} & \multicolumn{1}{c|}{{\color[HTML]{333333} \textbf{0.3705}}} & {\color[HTML]{333333} \textbf{0.6419}} & {\color[HTML]{333333} 0.4244}          & \multicolumn{1}{c|}{{\color[HTML]{333333} \textbf{0.3608}}} & {\color[HTML]{333333} \textbf{0.6329}} & {\color[HTML]{333333} 0.4383} & {\color[HTML]{333333} \textbf{0.3861}} \\ \hline
\multicolumn{13}{c}{Movie and Music}                                                                                                                                                                                                                                                                                                                                                                                                                                                                                                                                                     \\ \hline
\multicolumn{1}{c|}{topK}       & \multicolumn{6}{c|}{topK=5}                                                                                                                                                                                                                                                          & \multicolumn{6}{c}{topK=10}                                                                                                                                                                                                                                     \\ \hline
\multicolumn{1}{c|}{Domain}     & \multicolumn{3}{c|}{Movie}                                                                                                                    & \multicolumn{3}{c|}{Music}                                                                                                           & \multicolumn{3}{c|}{Movie}                                                                                                                    & \multicolumn{3}{c}{Music}                                                                                       \\ \hline
\multicolumn{1}{c|}{Metrics}    & HR                                     & NDCG                                   & \multicolumn{1}{c|}{MRR}                                    & HR                                     & NDCG                          & \multicolumn{1}{c|}{MRR}                                    & HR                                     & NDCG                                   & \multicolumn{1}{c|}{MRR}                                    & HR                                     & NDCG                          & MRR                                    \\
\multicolumn{1}{c|}{Uniform}    & 0.4831                                 & 0.3561                                 & \multicolumn{1}{c|}{0.3168}                                 & 0.5123                                 & 0.3875                        & \multicolumn{1}{c|}{0.3470}                                 & 0.6214                                 & 0.4005                                 & \multicolumn{1}{c|}{0.3350}                                 & 0.6382                                 & 0.4295                        & 0.3644                                 \\
\multicolumn{1}{c|}{Laplace}    & \textbf{0.4931}                        & \textbf{0.3684}                        & \multicolumn{1}{c|}{\textbf{0.3291}}                        & 0.5297                                 & \textbf{0.4063}               & \multicolumn{1}{c|}{\textbf{0.3660}}                        & \textbf{0.6274}                        & \textbf{0.4117}                        & \multicolumn{1}{c|}{\textbf{0.3464}}                        & \textbf{0.6578}                        & \textbf{0.4477}               & \textbf{0.3823}                        \\
\multicolumn{1}{c|}{MVGaussian} & 0.4734                                 & 0.3517                                 & \multicolumn{1}{c|}{0.3103}                                 & 0.4992                                 & 0.3824                        & \multicolumn{1}{c|}{0.3433}                                 & 0.6113                                 & 0.3949                                 & \multicolumn{1}{c|}{0.3281}                                 & 0.6294                                 & 0.4242                        & 0.3603                                 \\
\multicolumn{1}{c|}{Gaussian}   & {\color[HTML]{333333} 0.4891}          & {\color[HTML]{333333} 0.3632}          & \multicolumn{1}{c|}{{\color[HTML]{333333} 0.3224}}          & {\color[HTML]{333333} \textbf{0.5314}} & {\color[HTML]{333333} 0.4037} & \multicolumn{1}{c|}{{\color[HTML]{333333} 0.3653}}          & {\color[HTML]{333333} 0.6241}          & {\color[HTML]{333333} 0.4076}          & \multicolumn{1}{c|}{{\color[HTML]{333333} 0.3404}}          & {\color[HTML]{333333} 0.6550}          & {\color[HTML]{333333} 0.4442} & {\color[HTML]{333333} 0.3819}          \\ \hline
\multicolumn{13}{c}{Music and Book}                                                                                                                                                                                                                                                                                                                                                                                                                                                                                                                                                      \\ \hline
\multicolumn{1}{c|}{topK}       & \multicolumn{6}{c|}{topK=5}                                                                                                                                                                                                                                                          & \multicolumn{6}{c}{topK=10}                                                                                                                                                                                                                                     \\ \hline
\multicolumn{1}{c|}{Domain}     & \multicolumn{3}{c|}{Music}                                                                                                                    & \multicolumn{3}{c|}{Book}                                                                                                            & \multicolumn{3}{c|}{Music}                                                                                                                    & \multicolumn{3}{c}{Book}                                                                                        \\ \hline
\multicolumn{1}{c|}{Metrics}    & HR                                     & NDCG                                   & \multicolumn{1}{c|}{MRR}                                    & HR                                     & NDCG                          & \multicolumn{1}{c|}{MRR}                                    & HR                                     & NDCG                                   & \multicolumn{1}{c|}{MRR}                                    & HR                                     & NDCG                          & MRR                                    \\ \hline
\multicolumn{1}{c|}{Uniform}    & 0.4403                                 & 0.3305                                 & \multicolumn{1}{c|}{0.3054}                                 & 0.4394                                 & 0.3338                        & \multicolumn{1}{c|}{0.2989}                                 & 0.5629                                 & 0.3700                                 & \multicolumn{1}{c|}{0.3212}                                 & 0.5576                                 & 0.372                         & 0.3147                                 \\
\multicolumn{1}{c|}{Laplace}    & 0.4614                                 & 0.3568                                 & \multicolumn{1}{c|}{0.3230}                                 & \textbf{0.4623}                        & \textbf{0.3572}               & \multicolumn{1}{c|}{\textbf{0.3217}}                        & 0.5766                                 & 0.3954                                 & \multicolumn{1}{c|}{0.3388}                                 & \textbf{0.5768}                        & \textbf{0.3928}               & 0.3364                                 \\
\multicolumn{1}{c|}{MVGaussian} & 0.4392                                 & 0.3431                                 & \multicolumn{1}{c|}{0.3056}                                 & 0.4325                                 & 0.3303                        & \multicolumn{1}{c|}{0.2952}                                 & 0.5675                                 & 0.3818                                 & \multicolumn{1}{c|}{0.3219}                                 & 0.5590                                 & 0.3700                        & 0.3116                                 \\
\multicolumn{1}{c|}{Gaussian}   & {\color[HTML]{333333} \textbf{0.4686}} & {\color[HTML]{333333} \textbf{0.3683}} & \multicolumn{1}{c|}{{\color[HTML]{333333} \textbf{0.3282}}} & {\color[HTML]{333333} 0.4496}          & {\color[HTML]{333333} 0.3493} & \multicolumn{1}{c|}{{\color[HTML]{333333} 0.3155}}          & {\color[HTML]{333333} \textbf{0.5942}} & {\color[HTML]{333333} \textbf{0.4034}} & \multicolumn{1}{c|}{{\color[HTML]{333333} \textbf{0.3444}}} & {\color[HTML]{333333} 0.5669}          & {\color[HTML]{333333} 0.3865} & {\color[HTML]{333333} \textbf{0.3369}} \\ \hline
\end{tabular}
}}
\end{table*}
\subsubsection{\textbf{The Effects of Different Priors}}
\label{sec:diff_priors}
In Section~\ref{sec:objective_implementation}, we take standard Gaussian distributions as priors. Different priors have different prior knowledge for the learned user preferences. Thus it is interesting to see the effects of different priors.
In this part, we explore the effects of different priors on the model performance. In particular, we apply four common priors in generative modeling and show the results in Table~\ref{table:diff_priors}. From Table~\ref{table:diff_priors}, we have the following key observations:
\begin{itemize}
    \item Compared to other priors, uniform prior consistently leads to poor recommendation performance. It is because users usually have non-uniform interests on different topics and the uniform distribution does not match the implicit distribution of user preferences.
    \item Gaussian and Laplace prior have comparable performance. On Movie \& Music, Laplace prior gain better performance than Gaussian prior, while on other two benchmarks, they have slight gaps and Gaussian prior has better performance in some cases. Thus the choice of prior should be data-dependent.
    \item MVGaussian prior generally does not have better performance than Gaussian prior. This is mainly because it is hard to define the parameters of MVGaussian and it is easy to involve human bias. Instead, the most common prior (\emph{i.e.} Gaussian) can provide satisfied performance.
\end{itemize}
\begin{figure}[t]
\centering
\begin{minipage}[t]{0.45\textwidth}
\centering
\includegraphics[width=\textwidth]{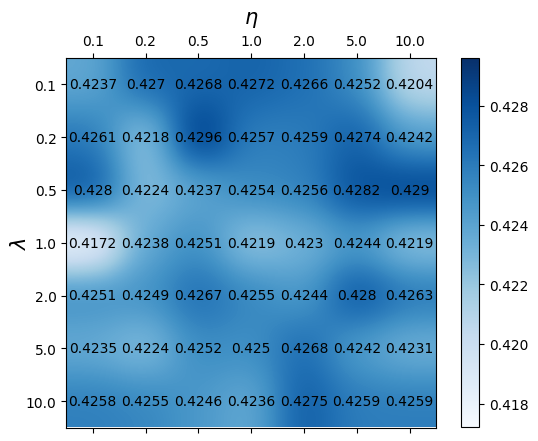}
\vspace{-20pt}
\captionsetup{justification=centering}
\caption*{(a) \footnotesize{Movie \& Book \\Movie-NDCG@10}}
\end{minipage} 
\begin{minipage}[t]{0.45\textwidth}
\centering
\includegraphics[width=\textwidth]{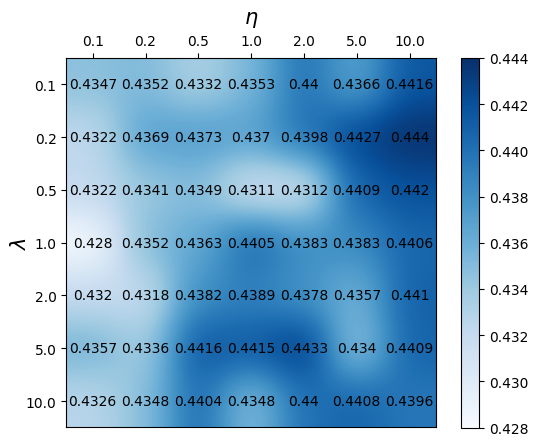}
\vspace{-20pt}
\captionsetup{justification=centering}
\caption*{(b) \footnotesize{Movie \& Book\\Book-NDCG@10}}
\end{minipage} \\
\begin{minipage}[t]{0.45\textwidth}
\centering
\includegraphics[width=\textwidth]{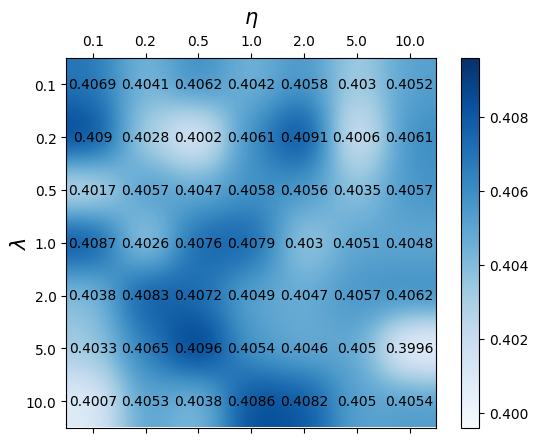}
\vspace{-20pt}
\captionsetup{justification=centering}
\caption*{(c) \footnotesize{Movie \& Music\\Movie-NDCG@10}}
\end{minipage} 
\begin{minipage}[t]{0.45\textwidth}
\centering
\includegraphics[width=\textwidth]{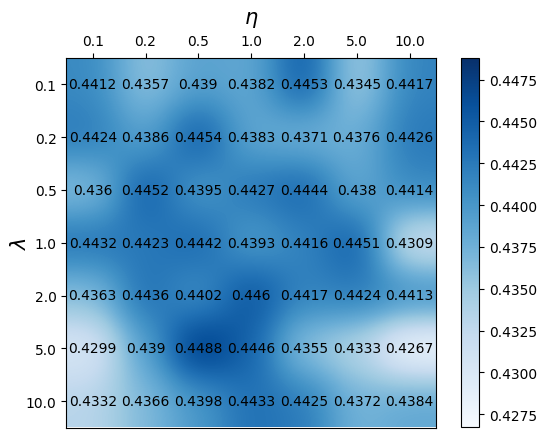}
\vspace{-20pt}
\captionsetup{justification=centering}
\caption*{(d) \footnotesize{Movie \& Music\\Music-NDCG@10}} 
\end{minipage} 
\caption{The effects of hyper-parameters $\lambda,\eta$ on two datasets. We show the results of two datasets here to illustrate the observations and the results of another dataset follow similar pattern.}
\label{figure:hyperparameter}
\end{figure}
\subsubsection{\textbf{Hyper-parameter Sensitivity}}
In ETL, $\lambda$ controls the weight of the equivalent transformation and $\eta$ weights the prior regularization. To investigate how the hyper-parameters $\lambda,\eta$ influence the performance of ETL, we conduct an experiment to study the sensitivity of these two hyper-parameters. The corresponding results are shown in Figure~\ref{figure:hyperparameter}. 

From this figure, we can see that the best hyper-parameter setting is different on different datasets. For example, in Figure~\ref{figure:hyperparameter} (b), ETL performs better when $\eta$ is around 10.0 and $\lambda$ is around 0.2 on Movie \& Book-Book. While for Movie \& Music-Music in Figure~\ref{figure:hyperparameter} (d), the hyper-parameters are around $\eta=0.5,\lambda=5.0$. This is because different datasets have different distributions and require different weights of the equivalent transformation and the prior regularization.


\section{Conclusion and Future Work}
In this paper, based on joint distribution matching, we propose a novel method called ETL which advocates to capture both the overlapped and domain-specific features for CDR. ETL takes a novel equivalent transformation assumption across domains for better user preference modeling and user behavior prediction.
Through extensive experiments, we find that only learning one kind of features is limited for the performance of a CDR model. 
A well-performed CDR model should capture the domain-specific features on the base of the well-captured overlapped features. The proposed ETL works in a principled way by maximizing the joint likelihoods of the joint user behaviors across domains, and does not require heuristic ways of choosing training users or designing the network architectures.

Although ETL has shown better performance than previous methods, there are still some inadequacies that limit its potential. Firstly, in this paper, we make a simplified Gaussian assumption for the joint prior $p(z_{x},z_{y})$ in Section~\ref{sec:PRL}.
A complex prior that provides more informative prior knowledge can be explored later.
Secondly, in common CDR, usually popular related domains are considered for recommendation. In the future, it is interesting to study how to do the recommendation among unpopular domains. Thirdly, how to incorporate ETL with auxiliary information to boost the recommendation performance and even solve the cold-start problem is also interesting.

\bibliographystyle{ACM-Reference-Format}
\bibliography{sample-base}

%









\end{document}